\documentclass[a4paper,fleqn,usenatbib]{mnras}
\usepackage{dcolumn}
\usepackage{bm}
\usepackage{lscape}
\usepackage{graphicx}
\usepackage{subfigure}
\usepackage{hyperref}
\usepackage[hyphenbreaks]{breakurl}
\usepackage{rotating}
\usepackage{tikz}
\usepackage{setspace}
\usepackage{array} 
\usepackage{adjustbox}
\usepackage{siunitx}
\usepackage{float}

\graphicspath{{Pics/}}

\begin{document}

\title[DEVILS: Mass Growth of Bulges and disks]{Deep Extragalactic VIsible Legacy Survey (DEVILS): The emergence of bulges and decline of disk growth since $z = 1$.
}

\author[Hashemizadeh et al.]{Abdolhosein Hashemizadeh$^{1}$\thanks{E-mail: abdolhosein.hashemizadeh@research.uwa.edu.au}, Simon P. Driver$^{1}$, Luke J.~M.~Davies$^1$,
	\newauthor
	Aaron S. G. Robotham$^1$, 
	Sabine Bellstedt$^1$, 
    Caroline Foster$^{2,3}$, 
	\newauthor
    Benne W. Holwerda$^4$, 
	Matt Jarvis$^{5,6}$,
    Steven Phillipps$^{7}$,
    Malgorzata Siudek$^{8,9}$, 
    \newauthor
    Jessica E. Thorne$^1$, 
    Rogier A. Windhorst$^{10}$, 
    Christian Wolf$^{11}$ \\   
	\\
$^1$ICRAR, The University of Western Australia, 35 Stirling Highway, Crawley WA 6009, Australia\\
$^2$Sydney Institute for Astronomy, School of Physics, A28, The University of Sydney, NSW, 2006, Australia \\
$^3$ARC Centre of Excellence for All Sky Astrophysics in 3 Dimensions (ASTRO 3D)\\
$^4$Department of Physics and Astronomy, University of Louisville, Natural Science Building 102, 40292 KY Louisville, USA \\
$^5$Department of Astrophysics, University of Oxford, The Denys Wilkinson Building, Keble Road, Oxford OX1 3RH, UK\\
$^6$Department of Physics, University of the Western Cape, Bellville 7535, South Africa \\
$^7$Astrophysics Group, School of Physics, University of Bristol, Tyndall Avenue, Bristol BS8 1TL, UK \\
$^8$Institut de F\'{\i}sica d'Altes Energies (IFAE), The Barcelona Institute of Science and Technology, 08193 Bellaterra (Barcelona), Spain \\
$^9$National Centre for Nuclear Research, ul. Pasteura 7, 02-093, Warsaw, Poland \\
$^{10}$School of Earth and Space Exploration, Arizona State University, Tempe, AZ 85287-1404 \\
$^{11}$Research School of Astronomy and Astrophysics, Australian National University, Cotter Road, Canberra, ACT 2611, Australia
}

\date{Accepted XXX. Received YYY; in original form ZZZ}

\pubyear{\the\year}


\label{firstpage}
\pagerange{\pageref{firstpage}--\pageref{lastpage}}
\maketitle

\begin{abstract}
We present a complete structural analysis of the ellipticals (E), diffuse bulges (dB), compact bulges (cB), and disks (D) within a redshift range $0 < z < 1$, and stellar mass $\log_{10}(\mathrm{M}_*/\mathrm{M}_\odot) \geq 9.5$ volume-limited sample drawn from the combined DEVILS and HST-COSMOS region. We use the {\sc ProFit} code to profile over $\sim35,000$ galaxies for which visual classification into single or double-component was predefined in Paper-I. Over this redshift range, we see a growth in the total stellar mass density (SMD) of a factor of 1.5. At all epochs we find that the dominant structure, contributing to the total SMD, is the disk, and holds a fairly constant share of $\sim60\%$ of the total SMD from $z = 0.8$ to $z = 0.2$, dropping to $\sim30\%$ at $z = 0.0$ (representing $\sim33\%$ decline in the total disk SMD). Other classes (E, dB, and cB) show steady growth in their numbers and integrated stellar mass densities. By number, the most dramatic change across the full mass range is in the growth of diffuse bulges. In terms of total SMD, the biggest gain is an increase in massive elliptical systems, rising from 20\% at $z = 0.8$ to equal that of disks at $z = 0.0$ (30\%) representing an absolute mass growth of a factor of 2.5. Overall we see a clear picture of the emergence and growth of all three classes of spheroids over the past 8 Gyrs, and infer that in the later half of the Universe's timeline spheroid forming-processes and pathways (secular evolution, mass-accretion, and mergers) appear to dominate mass transformation over quiescent disk growth.
\end{abstract}

\begin{keywords} galaxies: formation - galaxies: evolution - galaxies: bulges - galaxies: disk - galaxies: elliptical - galaxies: mass function - galaxies: structure - galaxies: general
\end{keywords}

\setlength{\extrarowheight}{0pt}

\section{Introduction}
\label{sec:intro}

Galaxies can experience significant morphological and structural transformation over cosmic time, from clumpy high redshift star-forming disks to smooth red spheroidal systems at the present day (e.g., \citealt{Trujillo07}; \citealt{vanDokkum10}; \citealt{Huertas-Company15}; \citealt{dosReis20}; \citealt{Hashemizadeh21}). However, there are still many open questions as to how galaxies build-up their stellar mass, how it is distributed to form the various structural components, and how these substructures evolve, resulting in the plethora of morphological types observed in the local Universe.  

Two-dimensional photometric decompositions of galaxies have been used in numerous studies to understand the formation pathways of different galaxy types. The earliest 2D decomposition endeavours came from \cite{Byun95}; \cite{Andredakis95} and \cite{deJong96}, which gave us our first understanding of the light distribution variation across different galaxy types. Historically, due to the computational complexity, single S\'ersic profiles (\citealt{Sersic63}) have been used for profile fitting of large sample of galaxies (e.g., \citealt{Simard02}; \citealt{Wuyts11} and \citealt{vanderWel12}; \citealt{Kelvin14}). Galaxies, however, are often more complex requiring extra components such as bulge, bar, etc., to be robustly fit. For example, a two-component model consisting of a spheroidal bulge (\citealt{deVaucouleurs48}) and an extended near exponential disk (\citealt{Freeman70}) have been used to great success in describing the light profile of galaxies (e.g., \citealt{Allen06}; \citealt{Simard11}; \citealt{Mendel14};\citealt{Salo15}; \citealt{Lange16}; \citealt{Dimauro18}; \citealt{Cook19}; \citealt{dosReis20}). Going further, several studies have developed kinematic structural decomposition methods using advanced IFU spectroscopy (e.g., \citealt{Emsellem07}; \citealt{Taranu17}). However, these have so far only been applied to relatively small samples of galaxies, $< 1000$, mostly at low redshifts due to the required high signal to noise ratio (e.g., \citealt{Johnston17}; \citealt{Zhu18}; \citealt{Tabor19}; \citealt{Zhu20}; and \citealt{Oh20}).  

The evolution of galaxies, and particularly multi-component systems, are inevitably tied to the disk and bulge formation scenarios. Currently two leading possible bulge-formation scenarios have been proposed. First, the ``early-bulge formation'' scenario predicts that mergers of small systems in the early Universe resulted in the formation of a spheroidal, pressure-supported system (e.g., \citealt{Aguerri01}; \citealt{Driver13}). Following this a disk grows around the bulge through various gas accretion events. A bulge that has formed in this manner is a compact structure known as a \textit{classical bulge} and is dynamically hot, featureless, and similar to a dry major merger remnant, an \textit{elliptical} galaxy (\citealt{Fisher08}). Second, the ``late-bulge formation'' scenario proposes that disks form first and then bulges form through in-situ events within the disk, such as disk instabilities and epicyclic motions (\citealt{Elmegreen08}). In this scenario, disk instabilities can lead to the flow of gas towards the centre of the gravitational potential and epicyclic motions amplify over time once the disk is stable, causing centralised star-formation and the growth of a bulge inside the already established disk. This type of bulge is traditionally known as a \textit{pseudo-bulge} \citep{Kormendy04}. Unlike classical-bulges, pseudo-bulges are dynamically cold and rotationally supported (\citealt{Kormendy93}; \citealt{Gao20}). In terms of colour, stellar population and metallicity, pseudo-bulges are more similar to the outer disk than classical-bulges or ellipticals (\citealt{Fisher06}; \citealt{Du20}; \citealt{Gao20}). Morphologically, pseudo-bulges and classical-bulges are argued to be distinguishable through their S\'ersic indices, with former having S\'ersic indices close to unity ($n \sim 1$), i.e., a near-exponential surface brightness profile, and latter having a higher S\'ersic index ($n > 2$)  more akin to that of spheroids (\citealt{Andredakis94}; \citealt{Andredakis95}; \citealt{Fisher06}; \citealt{Mendez-Abreu10}). However, recent kinematic decomposition studies find that S\'ersic index is not a good indicator of different types of bulge (\citealt{Krajnovic13}; \citealt{Zhu18}; \citealt{Schulze18}).  

A popular galaxy formation model called the two-phase scenario involves two periods of (i) a rapid high redshift in-situ star-formation at $2 < z < 6$ \citep{Oser10} and (ii) a successive phase dominated by minor mergers that are thought to form today's spheroidal structures (\citealt{Bluck12}; \citealt{McLure13}; \citealt{Robotham14}; \citealt{Ferreras17}; \citealt{Harmsen17}; \citealt{{DSouza18}}). Following this scenario, several studies compared the central surface brightness of massive high-$z$ spheroids with local galaxies and confirmed that they are structurally similar (\citealt{Hopkins09}; \citealt{Bezanson09}; \citealt{delaRosa16}). These studies reveal that high-$z$ ($z \simeq 1.5$) compact galaxies, also known as red nuggets \citep{Damjanov09}, are possibly at the centre of massive modern galaxies. While the \cite{Oser10} model mainly explains massive galaxies, more generally, by analysing the cosmic star-formation histories of disk galaxies and spheroids, \cite{Driver13} also proposed a two-phase galaxy evolution model. According to this model, compact bulges form first, and then from $z \approx 1.7$ disks grow around the bulges in low density environments and major mergers drive the formation of ellipticals in high-density environments. Note that in reality the above processes (mergers and disk instabilities) will both happen at all cosmic epochs but one process may dominate at high- or low-$z$.

By probing the dominant epochs of bulge and disk formation and the relative contribution of both pseudo- and classical-bulge in the galaxy population as a function of time, we can begin to disentangle their likely structural formation and evolution scenarios. While this is of paramount importance to our understanding of galaxy formation mechanisms, previous studies exploring the evolution of galaxy components on large evolutionary baselines have been hampered on a number of fronts. First, stellar populations cause colour gradients, so that measured parameters would vary due to bandpass shifting when comparing high-$z$ with low-$z$ images in the same wavelength band (e.g., \citealt{Kelvin14}; \citealt{Vulcani14}; \citealt{Kennedy16}). Second, dust is argued to distort our structural measurements including S\'ersic index and effective radius. Therefore, due to dust attenuation it is often impossible to measure the true profiles (e.g., \citealt{Pastrav13}). Third, galaxies are often more complicated than only a bulge+disk, so that it is not always obvious how to determine the appropriate number of components to fit (e.g., \citealt{Salo15}; \citealt{Lange16}). 

Motivated by this and recent software development in both source identification and structural fitting routines, we now revisit the structural decomposition of galaxies from $z \simeq 1$ to the present day.  We perform a robust 2D photometric decomposition of galaxies in the Deep Extragalactic VIsible Legacy Survey (DEVILS; \citealt{Davies18}) $10^h$ region (D10) using the Hubble Space Telescope (HST) imaging dataset of the Cosmic Evolution Survey (COSMOS). For our modeling, we make use of the state-of-the-art galaxy fitting software {\sc ProFit} (\citealt{Robotham17}). In this study, we adopt the perspective of fitting a disk and bulge complex, where the complex might be a diffuse-bulge, compact-bulge (dB, cB), and in some cases a combination. Using these decompositions, we explore the evolution of the stellar mass density contribution of structural components, and use this to propose a solution to the competing bulge-formation scenarios.

This work is structured as follows. Section \ref{sec:SampleSel} discusses the D10/ACS sample, in Section \ref{sec:ProfileFitting} we outline our fitting pipeline ({\sc GRAFit}) and the tools used therein, as well as the HST PSF modelling. The verification of our structural analysis as well as our method for distinguishing between dB and cB are described in Section \ref{sec:profselection} and we then explain the evolution of the SMF and SMD in Sections \ref{Sec:SMF_evol} and \ref{sec:rho}, respectively. Finally, we discuss and summarize our results in Section \ref{sec:discussion} and \ref{sec:summary}. 

Throughout this paper, we use a flat standard $\Lambda$CDM cosmology of $\Omega_{\mathrm{M}} = 0.3$, $\Omega_\Lambda = 0.7$ with $H_0 = 70 \mathrm{km}\mathrm{s}^{-1}\mathrm{Mpc}^{-1}$ (\citealt{Planck20}).
Magnitudes are given in the AB system \citep{Oke83}. 

\begin{figure*}
	\centering
	\includegraphics[width=0.85\linewidth]{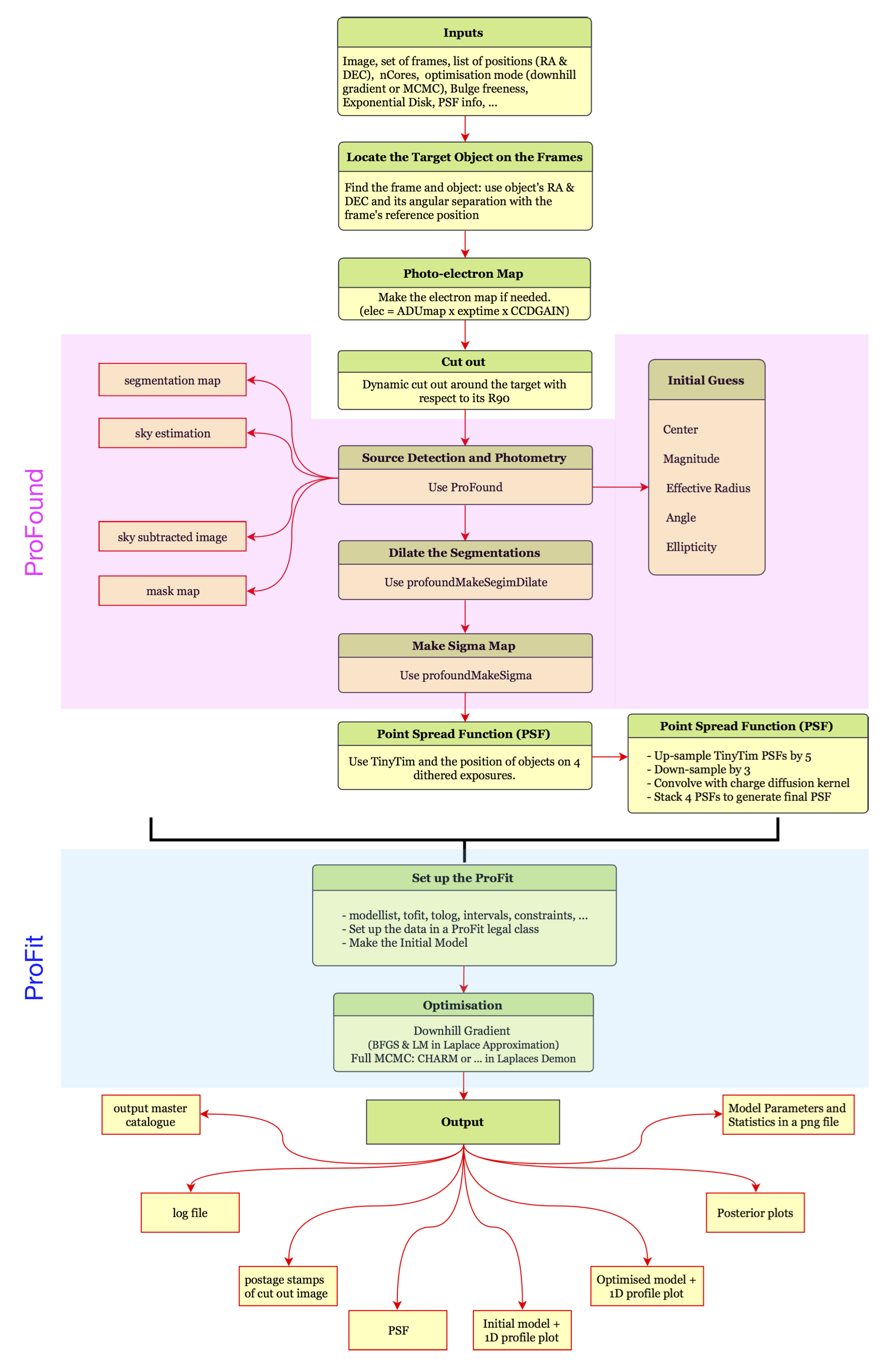}
	\caption{Flow diagram of {\sc GRAFit} containing five main parts; inputs and cut-out generation, running {\sc ProFound}, PSF generation; running {\sc ProFit} and outputs. }
	\label{fig:grafitflowdiagram}
\end{figure*}

\section{D10/ACS Sample and HST imaging data} 
\label{sec:SampleSel}

In this study, we use the D10/ACS sample constructed in \cite{Hashemizadeh21}, where we conducted a rigorous visual morphological classification of the sample into single- and double-component categories. This process initially made use of a number of automatic pre-classification methods followed by a full visual inspection by multiple classifiers. This sample classifies galaxies into double-component (bulge+disk; BD), pure-disk (D), elliptical (E), and compact (C) systems (see section 2 of the same paper). In this work, we combine `compacts' with `ellipticals' (E+C) as in the C subcategory is dominated by unresolved and, we believe, most likely compact spheroidal systems (see figure 21 in \citealt{Hashemizadeh21}). In brief, the D10/ACS sample we use here, was extracted from the $10^h$ the Cosmic Evolution Survey (COSMOS; \citealt{Scoville07}) region of the Deep Extragalactic VIsible Legacy Survey (DEVILS, \citealt{Davies18}). It consists of $35,803$ galaxies with multi-wavelength photometry from FUV to far-IR wavelengths (i.e., 0.2 to 500 micron; \citealt{Davies21}) and the sample extends up to $z = 1$ for systems with $\log_{10}(\mathrm{M}_*/\mathrm{M}_\odot) \geq 9.5$. We use a combination of photometric and spectroscopic (where available) redshifts as also described in \cite{Davies21}. The redshift and stellar mass limits were set in \cite{Hashemizadeh21} based on the limits to which our visual classifications can be considered reliable. This was established by visually inspecting galaxies drawn from the $\mathrm{M}_*-z$ plane and identifying the region where visual classification and 2D structural analysis was deemed viable by the three classifiers (SPD, LJMD, and SB, see section 1.3 of the same paper for more details on the sample completeness.

In order to perform our structural decomposition, we make use of the main imaging data of COSMOS, taken with the Advanced Camera for Surveys (ACS\footnote{ACS Hand Book: \href{http://www.stsci.edu/hst/acs/documents/handbooks/current/c05\_imaging7.html\#357803}{www.stsci.edu/hst/acs/documents/}}) on the Hubble Space Telescope (HST). It covers 1.7 square degrees centred at RA $150.12$ ($10:00:28.600$), and DEC $+2.21$ ($+02:12:21.00$) (J2000). The ACS observations used the F814W filter (I-band), providing good depth and flux measurements mostly red-ward of the $4000$\AA \, break out to $z = 1$, i.e., one is sampling red-ward of the Balmer and $4000$\AA \, break out to $z = 1$ at 814nm (\citealt{Hashemizadeh21}).
We use the drizzled COSMOS HST images for our bulge-disk decomposition analysis, which utilises the MultiDrizzle code \citep{Koekemoer03}. These data have been re-sampled to a pixel scale of $0.03$ arcsec from the original ACS pixel size of $0.05$ arcsec. 
The redshift and stellar masses used in the present work are taken from the DEVILS/D10 master redshift catalogue (\texttt{DEVILS\_D10MasterRedshiftCat\_v0.2}) and the \texttt{DEVILS\_D10ProSpectCat\_v0.3} catalogue, described in detail in \cite{Thorne20}.
For their stellar mass measurements, they perform SED fitting with the {\sc ProSpect} code \citep{Robotham20} and internally this adopts the \cite{BC03} stellar libraries, a \cite{Chabrier03} IMF, \cite{Charlot00} to model dust attenuation and \cite{Dale14} to model dust emission. \cite{Thorne20} uses the latest multiwavelength photometry measurements in the D10 field (\texttt{DEVILS\_PhotomCat\_v0.4}; see \citealt{Davies21}). They report stellar masses $\sim 0.2$ dex higher than in COSMOS2015 catalogue \citep{Laigle16} and this is traced to the inclusion of {\sc ProSpect}'s ability to fold in the evolving gas phase metallicity. See \cite{Thorne20} for full details.

\section{Profile Fitting} 
\label{sec:ProfileFitting}

In order to perform bulge-disk decompositions we need to consider a number of elements, which include: the pixels that are used for the fitting ({\sc ProFound}, \citealt{Robotham18}), the code for fitting the structural parameters ({\sc ProFit}, \citealt{Robotham17}), and our management of the end-to-end process including modelling of the {\it Hubble Space Telescope} point-spread function ({\sc GRAFit}). These are described below in full detail and the non-technical reader may wish to move forward to Section \ref{sec:profselection}, where we show and validate our resulting fits.

\subsection{ProFound}
Critical for a robust structural analysis is appropriate selection of the pixels used in the fitting. This process needs to ensure neighbouring objects are removed or flagged, but also aims to maximize the number of true pixels associated with the object. To achieve this we make use of {\sc ProFound} \citep{Robotham18}, an open source astronomical image analysis package. {\sc ProFound} analyses the image pixels, identifies all distinct sources, and provides a segmentation map for use in our fitting process. In addition, the code provides basic object size, and flux information that is used to define the initial parameters for the fitting code (this is non-essential but reduces the burn-in time of the MCMC fitting). 
The {\sc ProFound} segmentation map is a fundamental input for running {\sc ProFit} and specifies those pixels associated with the source, and from which the likelihood is computed. 
In addition to the segmentation, we also make use of {\sc ProFound}'s photometric measurements to provide initial estimates of the half-light radii, magnitudes, flux centers, axial ratios, and angle of rotation for the disk to be passed to {\sc ProFit}. 

Note that to determine initial parameters for the bulge and disk when fitting two-component systems, we choose to assign 20\% of the total flux to the bulge (i.e., B/T=0.2), and 20\% of the systemic angular size for the bulge $R_e$, the remaining flux was then assigned to the disk and the disk $R_e$ set to the systemic size. We reiterate that ProFit we specifically designed to overcome initial condition issues and hence we do not consider the starting conditions in any way critical to the fitting process.

The key distinction of {\sc ProFound} from previous source-detection codes is that it constructs segments that trace the outline of the galaxy as opposed to circles or ellipses. This is critical, as galaxies are not perfect ellipses, and elliptical apertures will not always accurately represent their flux distribution. Moreover, in complex regions, ellipses of neighbouring objects may overlap or intersect and disentangling the flux is complex. {\sc ProFound}'s solution is to define segments, based on the outer isophote, and to dilate these segments until they contain $95\%$ of the source's flux, essentially performing a curve of growth analysis. Notably, the dilation process does not allow segments to ever overlap and, therefore, each pixel is allocated entirely to a single object or left unallocated. This avoids the need to disentangle flux from objects, but can include some intervening light from neighbouring sources. On the whole, the dilation process is more aggressive for more luminous objects, and so pixels should end up assigned to the object that dominates the light. This aspect is somewhat of a trade-off between the errors associated with the dominant flux versus allowing for some cross-contamination. In our analysis we take the decision that the latter is less liable to gross error.

\subsection{ProFit}
\label{subsec:ProFit}

To determine bulge-disk decompositions, we use the Profile Fitting package, {\sc ProFit} (\citealt{Robotham17}). This was specifically designed for 2D structural analysis and can use a wide range of minimisation algorithms, essentially any of those available in {\sc R}, to obtain reliable solutions with robust error analysis, which are independent of the initial parameters. {\sc ProFit} and the low-level C++ library ({\sc libprofit}) are combined with a high-level R interface. Several profiles are in-built in {\sc ProFit} and any combination, as well as user defined profiles, can be used to model galaxy images. The in-built profiles are: S\'ersic, Core-S\'ersic, broken-exponential, Ferrer, Moffat, empirical King, point-source, and sky. We use a S\'ersic profile for both the disk and bulge components, i.e., a double S\'ersic fit.

This profile is described in \cite{Sersic63}; (also see \citealt{Graham05}) and provides an analytic formula for the light intensity profile as a function of radius:

\begin{equation}
I(r)=I_e \exp\bigg[-b_n \bigg(\left(\frac{r}{r_e}\right)^{1/n} - 1\bigg)\bigg],
\end{equation}

\noindent where $r_e$ is the effective radius, the radius containing half of the total flux, $I_e$ is the intensity at that radius and $n$ is known as the S\'ersic index that specifies the shape of the profile. For example, $n=0.5$, $n=1$ and $n=4$ represent Gaussian, exponential and de Vaucouleurs profiles (\citealt{deVaucouleurs48}), respectively. In general, it has been shown that disks are likely to follow an exponential profile, as opposed to spheroidal structures which tend to follow a near-de Vaucouleurs profile (e.g., \citealt{Patterson40}; \citealt{deVaucouleurs59}; \citealt{Freeman70}; \citealt{Kormendy77}).

Compared to {\sc GALFIT} (\citealt{Peng02,Peng10}), {\sc ProFit} is more robust to the effects of local minima due to its compatibility with several optimization algorithms such as Markov Chain Monte Carlo (MCMC); as was shown in \cite{Robotham17}. In this work, we use the Componentwise Hit-And-Run Metropolis (CHARM) algorithm in our MCMC sampling. We refer the reader to \cite{Robotham17} for further details regarding {\sc ProFit}.

\subsection{Pipeline: {\sc GRAFit}}
\label{sec:GRAFit}
In order to manage the full end-to-end process, including HST point-spread function measurement at the location of each galaxy, we developed an automatic galaxy decomposition pipeline, {\sc GRAFit}. {\sc GRAFit} is a series of modules and functions developed in {\sc R} with calls to {\sc ProFit}, {\sc ProFound} and other astronomical tools. {\sc GRAFit} is principally designed to operate on HST ACS data, however it can be used with any imaging survey. The full process is reasonably complex and hence the flow diagram for {\sc GRAFit} is shown in Figure \ref{fig:grafitflowdiagram}.

The minimum requirement to run {\sc GRAFit} is either a galaxy image in standard format (e.g., a FITS file) and a list of RA \& DEC positions indicating the location of the objects to be profiled, or a directory of pre-cutout postage-stamp images. In the case of the latter, {\sc GRAFit} identifies the correct image with which to extract the target object(s).   
By default, {\sc GRAFit} allocates both a bulge and a disk to the galaxy by performing double S\'ersic modelling, which distributes the total flux into bulge and disk. However, by altering the \texttt{nComp} flag, the user can also model a single S\'ersic profile. Since {\sc GRAFit} is efficiently programmed as parallel code, one can spread the tasks over multiple cores/nodes using the flags \texttt{nCores} and \texttt{ThreadMode}. It is hence supercomputer friendly, and has now been actively used on a number of supercomputer architectures. There are some other additional parameters that can be added (see Figure \ref{fig:grafitflowdiagram}).

{\sc GRAFit} is a modular-based script with a central master script, \texttt{GRAFitMaster}, that calls other modules internally. At the very first step, {\sc GRAFit} locates the object(s) by searching all the frames, runs {\sc ProFound}, identifies the segment associated with the desired object (position matching) and then makes a dynamic cutout around the galaxy. See Section \ref{subsec:Sky est} for more details. Initial estimation of the structural parameters are made, and a sigma (noise) map generated that indicates the errors in pixels across the image using \texttt{profoundMakeSigma}. This noise map includes a pixel-by-pixel mapping of the combined (in quadrature) sky noise (skyRMS), read noise and the RMS of the dark current noise, where pixels associated with interloping objects are masked out.
See Appendix \ref{app:GRAFit_more} for more details about {\sc GRAFit}.

\begin{figure}
	\centering
	\includegraphics[width=\columnwidth]{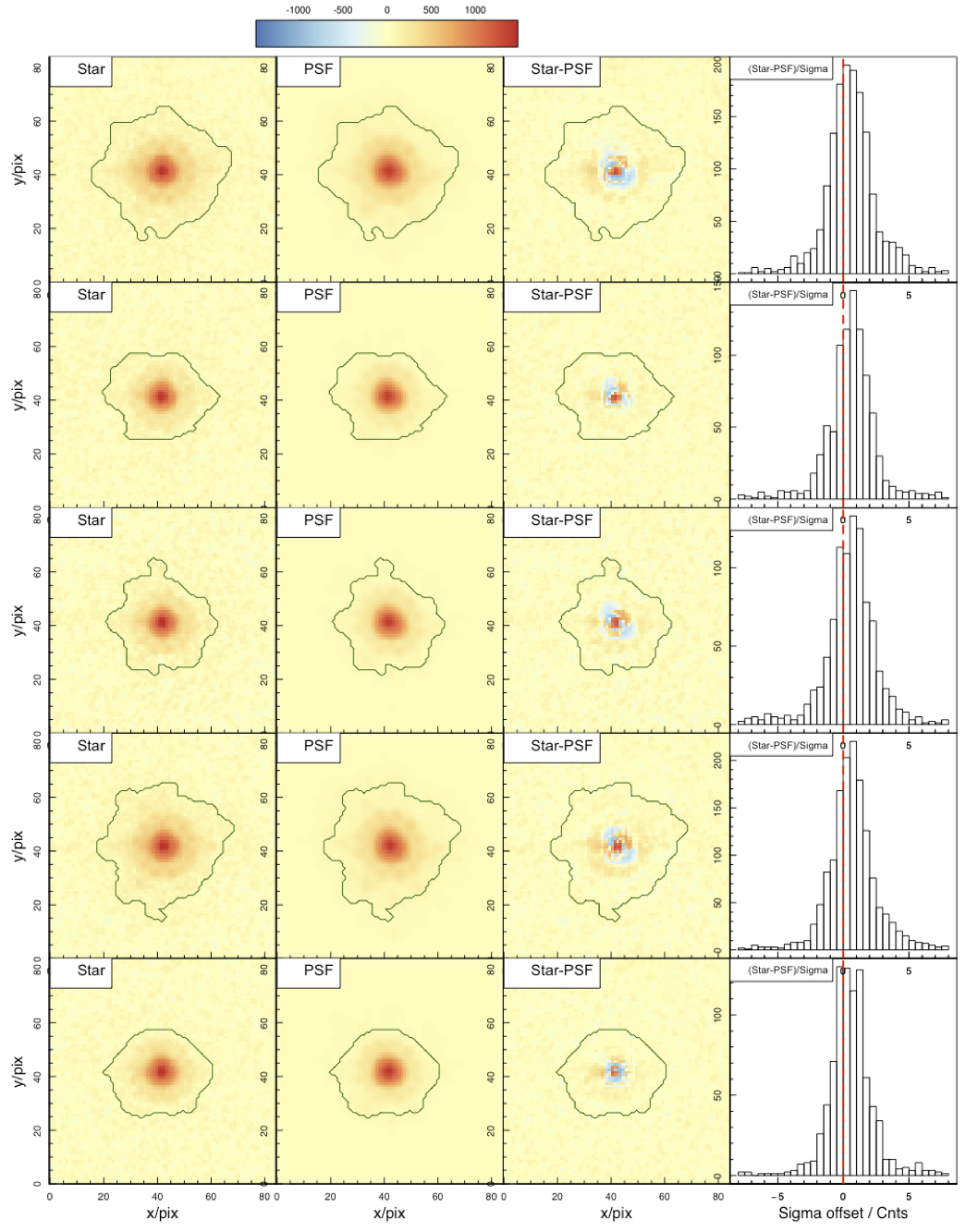}
	\caption{Five stars that were selected to subtract from associated PSFs (see the text for selection method). First column shows the stars in the drizzled images. Second and third columns display the PSF and the residual (Star-PSF), respectively. Fourth column represents the distribution of residual pixel values (Star-PSF/Sigma). }
	\label{fig:starsub_5star}
\end{figure}

\subsubsection{Modelling the HST/ACS PSF} 
\label{sec:PSF}

With the raw pixel cutout, the associated segment map, the initial parameter guess, and the noise map all prepared, the final -- and perhaps most complex -- aspect of the decomposition process is the {\it Hubble Space Telescope} ACS point spread function (PSF) modelling. Having an accurate PSF is obviously critical for modelling the central structures of galaxies. However, due to the off-axis location of the HST's ACS optics, the HST PSF is geometrically distorted and asymmetric along both X and Y directions \citep{Anderson06}. For this reason, we use the publicly available software {\sc Tiny Tim} \citep{Krist11} to model the PSF for different pixel positions on the ACS detections as observed through the F814W filter and also implement the function \texttt{tiny3} to apply the final ACS PSF geometrical distortion. 

In the mosaiced COSMOS HST/ACS imaging data, each pointing has been constructed with 4 distinct exposures (each 507 seconds), dithered by a few tens of pixels in both X and Y directions, to allow cosmic ray and bad pixel rejection. The dithering also compensates for the gap between the two ACS chips. We therefore revert to the four raw exposure frames to locate the position of our target object on each exposure. This enables us to generate four PSFs that we combine (i.e., stack) to produce the representative PSF for each object at 0.05 arcsec resolution. The mosaiced COSMOS HST/ACS imaging data is ultimately provided re-sampled to a pixel size of 0.03 arcsec (\citealt{Koekemoer07}). We therefore also re-sample the final stacked PSF to a 0.03 arcsec pixel scale (1.6 factor). {\sc Tiny Tim} only allows integer sub-sampling, so we first up-sample the PSF by a factor of 5 in the final stage of the {\sc Tiny Tim} process (by selecting SUB=5 in the \texttt{tiny3} function). We then down-sample the output PSF by a factor of 3 in an external step, leading us to the desired pixel size (0.03 arcsec/pix). 

{\sc Tiny Tim} does not automatically convolve the sub-sampled PSF with the CCD charge diffusion kernel. This is required, as point sources experience a slight blurring due to the charge diffusion into adjacent pixels. This reduces the sharpness of the PSF and causes a $\sim 0.5$ magnitude loss in WFC imaging at short wavelengths. Such blurring, which is also known as the pixel response function (PRF), is also field dependent due to the non-constant CCD thickness (12 to 17 microns for the WFC). See the ACS handbook\footnote{ACS Hand Book: \href{http://www.stsci.edu/hst/acs/documents/handbooks/current/c05\_imaging7.html\#357803}{www.stsci.edu/hst/acs/documents/}} 
for more detail. To simulate this blurring effect, {\sc Tiny Tim} provides the charge diffusion kernel as a $3\times3$ matrix in the PSF's header. This kernel is specific to the PSF's location and we use this kernel matrix and convolve it with our final re-sampled PSF. 

\subsubsection{Testing the HST PSF modelling} 
\label{sec:PSF_test}

To evaluate the accuracy of our PSF modelling, we perform a star subtraction test using the HST/COSMOS images. For this, we randomly select 5 bright unsaturated stars with half light radii of $R50\sim0.07$ arcsec (the typical radius seen), and with axial ratio of $> 0.9$ to ensure that the stars are unlikely to be binary systems. Note that R50 is obtained from our {\sc ProFound} analysis. See Appendix \ref{sec:star_sel} for more details on our star selection.

In Figure \ref{fig:starsub_5star}, the first column shows the star as observed (with segment boundary), the second column shows our modeled PSF to the same scale, and the third shows the residual having subtracted the PSF from the star. The rightmost column is the distribution of the pixel residual. Note that when subtracting the PSFs from stars, their centers must be accurately matched to the sub-pixel level to guarantee that there is no offset between the centres of the star and the PSF. We use {\sc ProFit} to interpolate the flux and find the sub-pixel center. For this we model a point source with the magnitude of the real star and convolve it with the PSF. We then run an optimization with the BFGS \footnote{Broyden-Fletcher-Goldfarb-Shanno} algorithm \citep{Broyden70} to find the accurate sub-pixel center and magnitude, and perform star subtraction precisely. We then analyze the residuals and the goodness of fits (GOF) calculated as GOF $=$ (PSF$-$star)/star and find $\mathrm{GOF} \sim 80\%$ implying that our PSFs simulate on average $\sim 80\%$ of the real stars' pixels with the most significant residual evident for the central pixel. 

For an additional quality check we apply a similar process to a star in the raw exposure frames. This is necessary to demonstrate that our PSF generation process such as re-sampling and charge diffusion kernel convolution is not affecting the PSF's profile, particularly for the central pixels. We present the result of this test in Figure \ref{fig:starsub_raw}. Here, the PSF is not required to be re-sampled as it is already matched with the original pixel scale of $0.05$ arcsec identical to the raw ACS imaging data. Again, a residual can be seen at the centre and the spread is in agreement with our previous conclusion. We therefore note that while {\sc Tiny Tim} represents the best model of the HST PSF, it comes with limitations. An aspect of the Hubble Space Telescope PSF not accounted for is the periodic ``breathing'' of the telescope referring to the small changes of the telescope's focus due to micron-scale movements of the secondary mirror \citep{Hasan94}. Currently this is outside the bounds of {\sc Tiny Tim} to model, and would require shifting to an empirical database of PSFs, currently under development at Space Telescope Science Institute (STSci). 

\section{One-component or two-component profile selection} 
\label{sec:profselection}

The GRAFit package produces viable outputs and three different models for all $\sim 35$k systems with only 33 cases ($< 0.07 \%$ of the sample) failing due to exceeding the computation (wall) time. We now need to determine which of our three fits is the most appropriate representation for each galaxy. 
As reported in \cite{Hashemizadeh21}, we explored the prospect of using different methods including cross matching with other available morphological catalogues to try to determine whether a galaxy contained a single dominant component or two distinguishable components.
Ultimately, we found no suitable solution that aligned well with our visual classifications. 
For this reason, we select either a S\'ersic (1C) or S\'ersic+S\'ersic (2C) profile based on our prior visual classifications. For elliptical (E) and pure-disk systems (D), we adopt the 1C profile, and for bulge+disk systems (BD) we adopt a 2C model. 
In a small fraction of cases 2C fits were poor due to an un-physical fit (e.g., $\mathrm{R}_{e, bulge} \gg \mathrm{R}_{e, disk}$). For these objects, we assess whether a S\'ersic+exponential disk profile solves the problem and find that for $\sim 3\%$ of the sample (1,072 objects) this profile describes the light distribution better than S\'ersic+S\'ersic. The rest of un-physical fits are flagged as poorly fitted in the final catalogue ($\sim 5\%$ of the full sample). This resulted in $3,812$ 1C elliptical systems ($\sim 11\%$), $15,608$ 2C two-component systems ($\sim 45\%$), $12,882$ 1C pure-disk systems ($\sim 37\%$), and $2,615$ unclassifiable systems ($\sim7\%$; representing objects visually identified as hard -interacting and visually disturbed systems etc.- or compact, or the aforementioned failed fits). Our fractions are to first order consistent with those \cite{Cook19} found for their xGASS sample. Note that compact systems are generally low-angular-sized spherical-like systems for which resolving their structures even with HST can be highly uncertain or in many cases impossible, see \cite{Hashemizadeh21} for more details.

\begin{figure*}
	\centering
	\includegraphics[width = \textwidth, angle = 0]{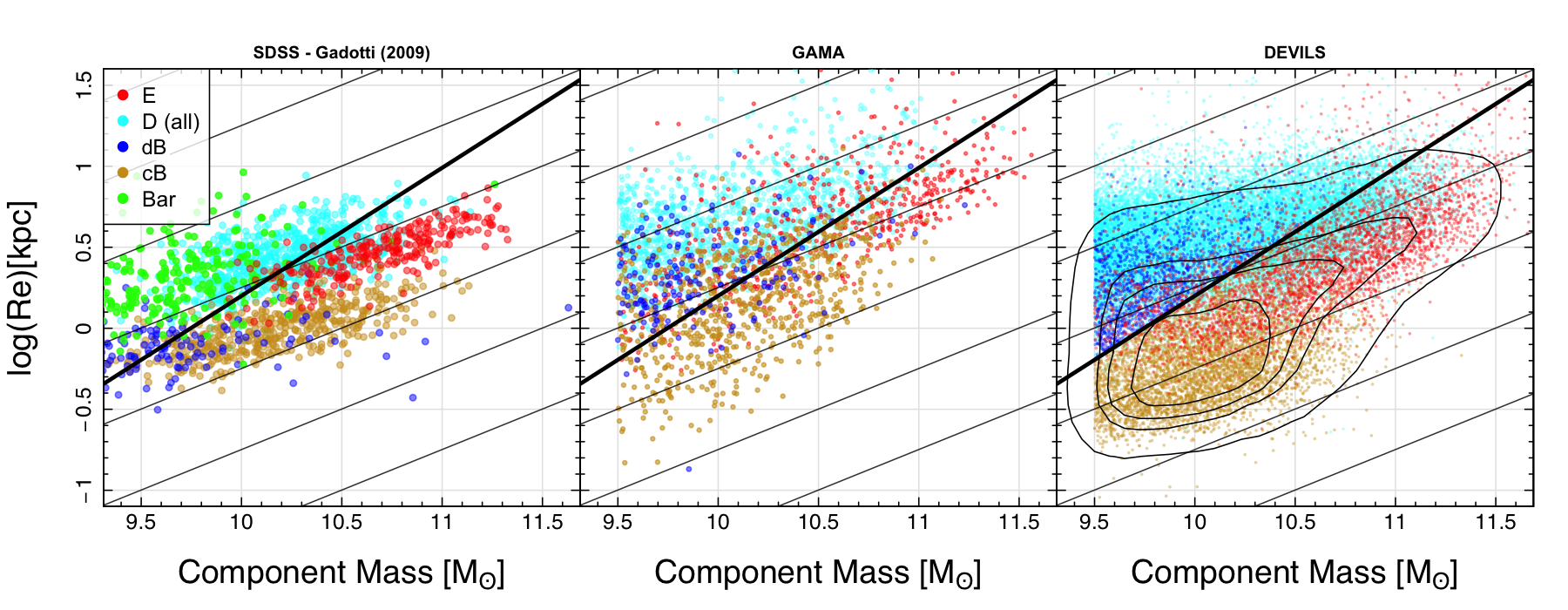}
	\caption{The size-mass plane for SDSS galaxies (left panel; \protect\citealt{Gadotti09}), GAMA galaxies (middle panel) and D10/ACS sample (right panel; this work). For completeness, we show the position of bars in SDSS galaxies in the left panel, although bars are not considered in GAMA and DEVILS sample. In the middle panel, the data is color coded based on our visual bulge classification of GAMA local galaxies \citep{Driver22}. Black solid lines correspond to our dB-cB separation line; following the equation $\mathrm{log}(\rm{R}_e/\mathrm{kpc}) = 0.79 \mathrm{log}(M_*/M_\odot) - 7.7$. Faint gray lines indicate constant stellar mass densities equivalent to $\mathrm{log}(\Sigma) = 11, 10, 9, 8, 7, 6, 5$, from top to bottom. }
	\label{fig:Bulge_M_Re}
\end{figure*}

\subsection{Distinguishing between diffuse- and compact-bulges} 
\label{sec:dB_cB_dist}

As a final step, we now attempt to separate our bulge components into ``compact''- and ``diffuse''-bulges (cB and dB, respectively), as they likely have different formation and evolutionary histories. However, we highlight that this distinction is problematic and increasingly challenging. This classification would be optimally done with kinematic data, but such large sample of kinematic data do not yet exist, especially not at these redshift ranges.  

Many studies have shown, by photometric and/or kinematic means, that the central regions of disk galaxies are often occupied by two types of structures (pseudo- and/or classical-bulges), see, e.g., the review of \cite{Kormendy04} where they highlight the differences between these structures. The definition of a pseudo-bulge is varied within the literature, and often depends on the information at hand, which can vary from a single-band image to full kinematic analysis. Here, for clarity, we elect to use the less-charged terminology of 'diffuse' and 'compact' which emphasises that in our case our distinction is based purely on visual classification criteria. In due course, and through further studies involving kinematic information, it may become clearer whether these classes do or do not equate to the kinematically distinct 'pseudo` and 'classical` bulges. To be clear our definition is therefore:

~

\noindent
$\bullet$ {\bf compact-bulge (cB)}: a high-stellar density, compact system, with no visible dust-lanes, asymmetries or distortions.

~

\noindent
$\bullet$ {\bf diffuse-bulge (dB)}: a low-stellar density, diffuse and extended system which may contain dust lanes and asymmetries.

~

\begin{figure*}
\begin{subfigure} 
  \centering
  \includegraphics[width=.49\linewidth]{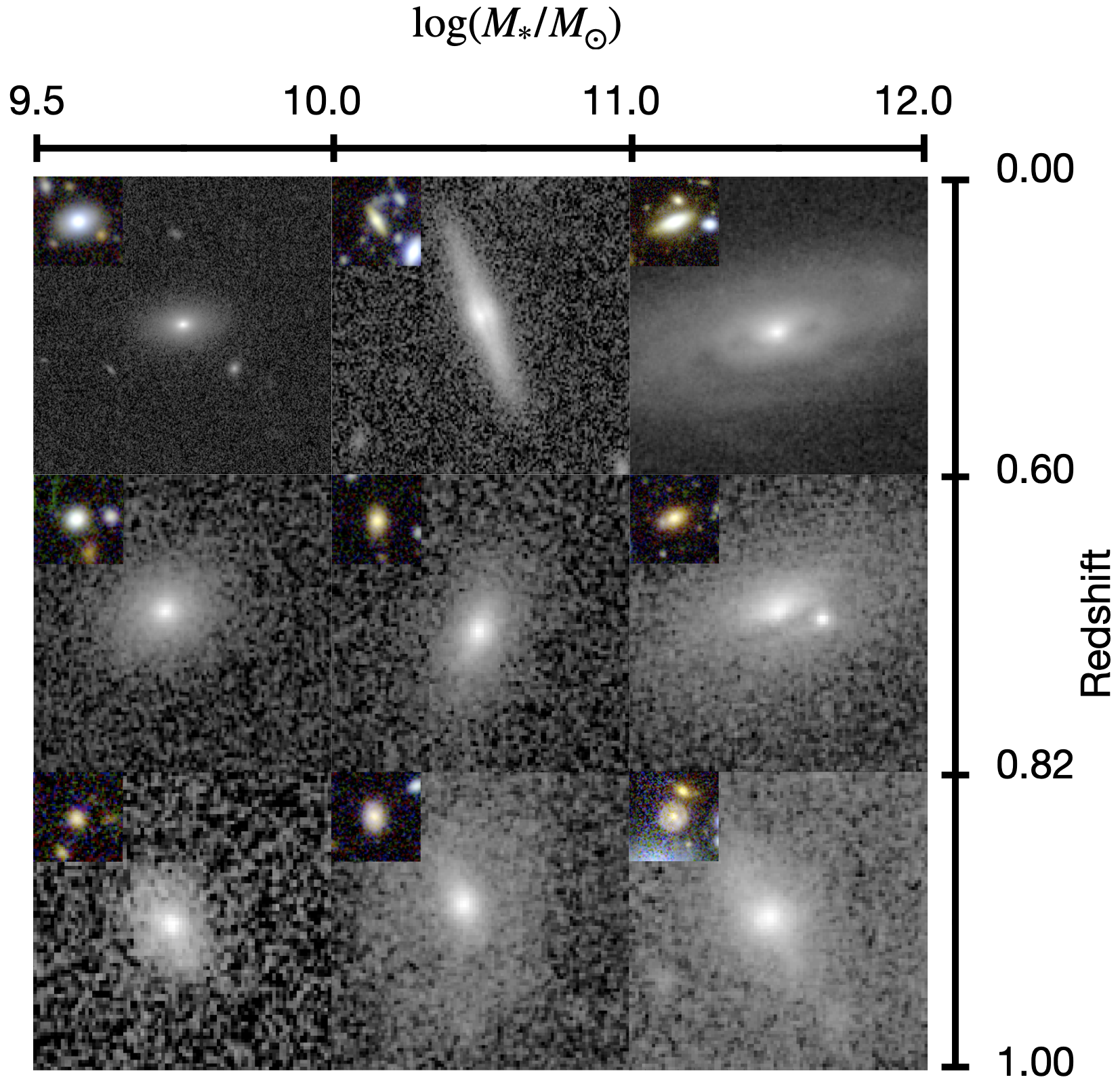}

\end{subfigure}
\begin{subfigure}
  \centering
  \includegraphics[width=.49\linewidth]{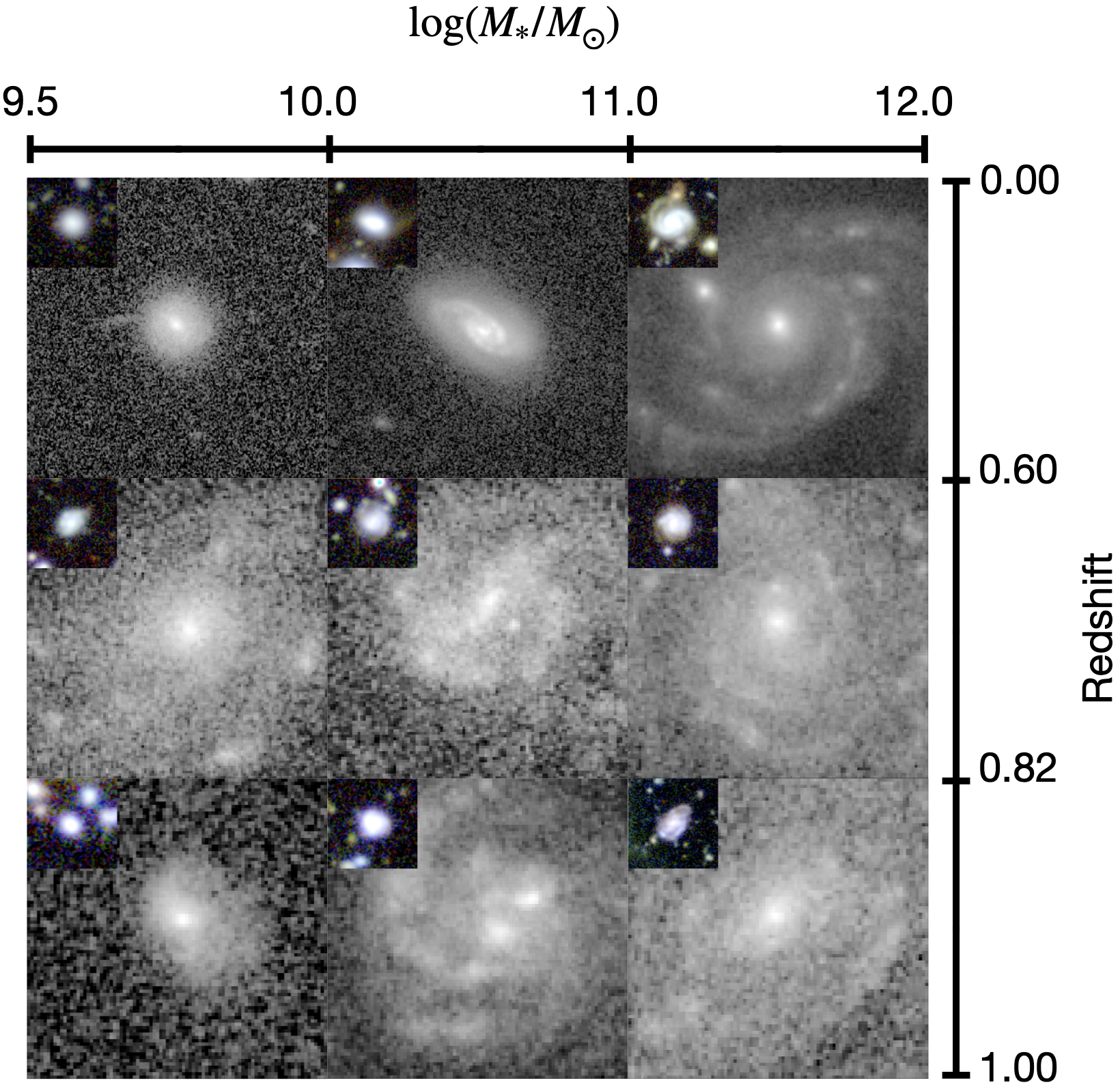}
\end{subfigure}
\caption{Random sample of galaxies harbouring a cB (left) and dB (right) as a function of redshift and stellar mass. The main image is cutout of galaxies in the ACS F814W filter while the inset color image is the SUBARU SuprimeCam gri combined image.}
\label{fig:dB_cB_stamps}
\end{figure*}

In our definition, a dB may therefore include the combination of a number of secondary perturbations including bars, rings and extended planar orbits that we are here aggregating into a central combination of structures. The motivation for this is to map this classification to our two-component fitting approach, as we do not believe fitting with additional components is viable or stable at this level of signal-to-noise and spatial resolution. Following this definition, we now explore the mass-size ($M_*-\rm{R}_e$) plane, which matches directly to the stellar surface density, and allows us to be guided by the visual classifications of dB/cB made for GAMA and pseudo-/classical-bulge for SDSS (based on the Kormendy relation; \citealt{Gadotti09}) galaxies at $z = 0$, as well as the distribution of our structural measurements, to attempt to select dB and cB structures in our sample. This method directly takes the bulge's stellar mass and effective radius into account rather than calculating the mean effective surface brightness within the effective radius ($<\mu_e>$) as in the Kormendy relation. This is expected to reduce propagation of uncertainties in $\rm{R}_e$ and $M_*$ into the calculation of the stellar surface density.

Note that we elect not to use a simple S\'ersic cut to separate dBs and cBs, as others have advocated, for a number of reasons. First, the bulge component S\'ersic index is fairly unstable (see Figure \ref{fig:sSSFR_Mass}), particularly given the uncertainty around the HST ACS PSF due to HST's ``breathing''. Second, dust can lower S\'ersic indices and also make the bulge appear larger (see e.g., \citealt{Pastrav13}). Our sample spans a broad redshift range where galaxies are also likely to become more dusty at higher-redshift (due to bandpass shifting, and higher star-formation rates). The fraction of massive galaxies with a dust-lane in the COSMOS region out to $z \sim 0.8$ is reported to be 80\% (\citealt{Sheth08}; \citealt{Holwerda12}). Third, the S\'ersic index is known to be wavelength dependent \citep{Kelvin14}, and hence a simple cut in a ``direct observable'' could introduce a redshift bias (due to bandpass shifting).
Fourth, a number of studies (e.g., \citealt{Gadotti09}; \citealt{Fisher16}) have shown that S\'ersic indices of pseudo- and classical-bulges overlap, when selected either visually (e.g., \citealt{Fisher08}), or via mean surface brightness (e.g., \citealt{Gadotti09}; using the Kormendy relation). This effect is also observed in our sample.

Finally, more recent results from the kinematic decomposition of disk galaxies with IFU observations have found no significant correlation between photometric S\'ersic index and kinematic properties (see e.g., \citealt{Krajnovic13}; \citealt{Zhu18}; \citealt{Schulze18}).  

Figure \ref{fig:Bulge_M_Re} compares the $M_*-\rm{R}_e$ relation for elliptical (E: red), Disks (D: cyan; representing both pure disk systems and disk components) dBs (blue) and cBs (gold) for our D10/ACS galaxies (right panel) with those drawn from the local SDSS (left panel) and GAMA (middle panel) surveys. We also show the bar component (green) from \cite{Gadotti09} work of the SDSS galaxies. This Figure highlights bulge classification of SDSS galaxies based on the \cite{Kormendy77} relation (left panel; \citealt{Gadotti09}) and GAMA galaxies classified through our visual inspections (middle panel). In the right panel, we show our D10/ACS sample. 

Note that lacking high-resolution colour information, we estimate the stellar mass of components by using our F814W bulge-to-total flux ratio (B/T), i.e., $M_*^\mathrm{Bulge} = \mathrm{B/T} \times M_*^\mathrm{Total}$ and $M_*^\mathrm{disk} = (1-\mathrm{B/T}) \times M_*^\mathrm{Total}$. The caveat here is that if the stellar population of the two components are different, then one can expect that the M/L are different, introducing errors into this method. However, this effect is unlikely to impact our results at the population scale.  

\begin{figure*}
	\centering
	\includegraphics[width = \textwidth, angle = 0]{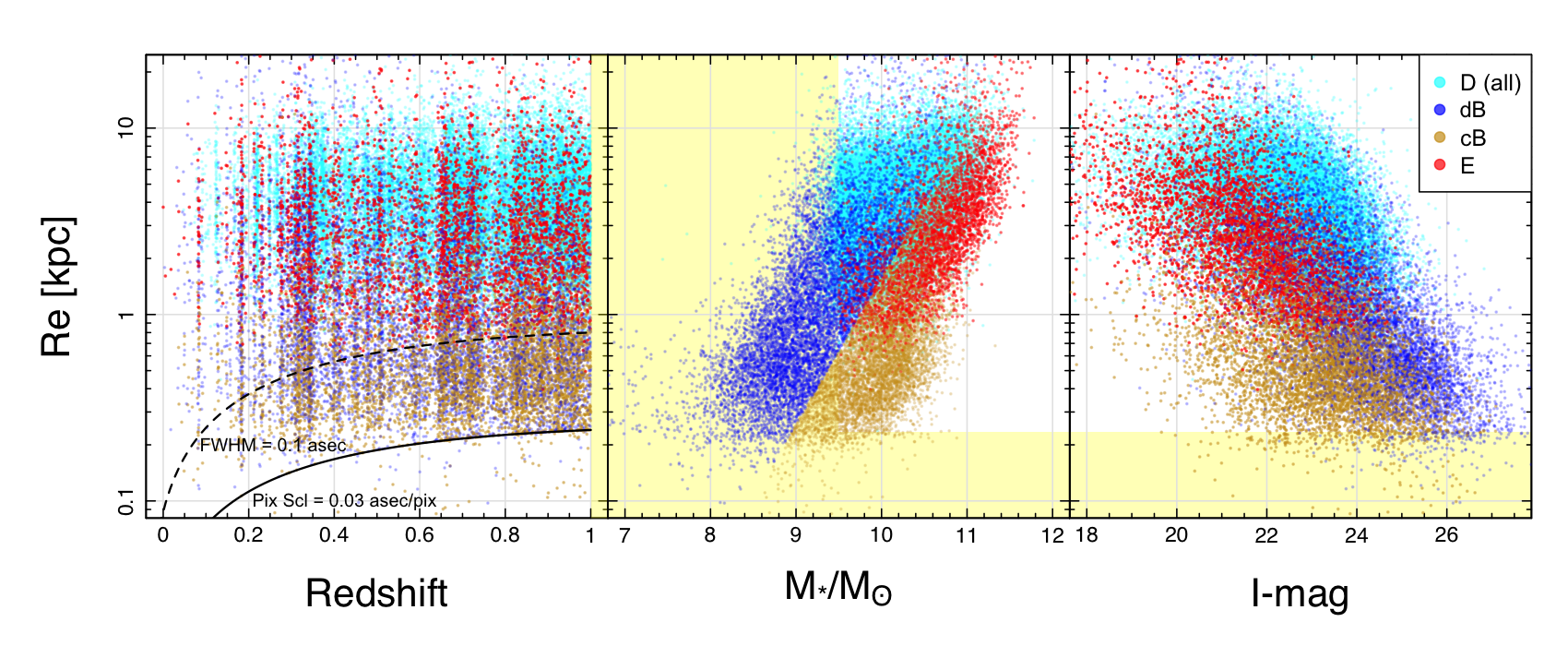}
    \caption{The relation between the effective radius, $\rm{R}_e$, of disks (cyan; including both pure disk systems and disk components), dB and cB (blue and gold) as well as ellipticals (red) with redshift, components' stellar mass and I-mag} (left, middle and right panels, respectively). The curve in the left panel represents our imaging pixel scale ($0.03$ arcsec per pixel) converted into physical size. Vertical yellow boundaries in the left and middle panels represent our redshift and stellar mass limit, respectively. Horizontal yellow boundaries represent the completeness of our data in size, i.e., $\rm{R}_e$ = 0.235 kpc and stellar mass, i.e., $M_* > 10^{9.5} M_\odot$.
	\label{fig:B_D_scatterPlot}
\end{figure*}

Figure \ref{fig:Bulge_M_Re} indicates that in the GAMA and SDSS data (left and middle panels) we see, despite obvious intermingling, a relatively clear demarcation between dB and cB. In the DEVILS data (right panel), we see a clumped population, which we identify as cBs (objects with higher stellar surface density), and a more dispersed population which we identify as dB (as one might expect from an amalgam of central perturbations, following our definition). We identify the line given by $\mathrm{log}(\rm{R}_e/\mathrm{kpc}) = 0.79 \mathrm{log}(M_*/M_\odot) - 7.7$, as providing a good demarcation across all three panels (surveys), and this is shown as the black lines on Figure \ref{fig:Bulge_M_Re} (essentially a cut slightly offset from a line of constant surface stellar mass density, shown as grey lines). Note that a fuller investigation of the mass-size relation will be provided in a forthcoming DEVILS paper.

Note that we show our visual dB/cB separation of GAMA galaxies in Figure \ref{fig:Bulge_M_Re} to highlight how our dB/cB separation line is guided by this data. However, going forward we will now consistently use the same dB/cB identification using the above cut for both the GAMA and the D10/ACS data. They possibly suffer from different systematic errors, e.g., how PSFs are made and how stable they are etc.

Given this distinction between the two bulge structures we find the majority of bulges in the Universe, by number, to be dB. Overall, 58\% of our double component galaxies contain a dB, while 42\% of them have a cB. However, when we only consider components above our imposed stellar mass limit of $\mathrm{log}(M_*/M_\odot) > 9.5$, as we show in Figure \ref{fig:Bulge_M_Re}, we find that dBs and cBs constitute 31\% and 69\% of bulges, respectively. 

Finally, Figure \ref{fig:dB_cB_stamps} display a random set of our galaxies classified as cBD (compact-Bulge+Disk) and dBD (diffuse-Bulge+Disk), respectively, in regular bins of stellar mass and redshift. The Figures indicate that dBs typically lie in bluer, more star-forming systems than cBs, with their outer disks displaying more structure, i.e., spiral arms, star-formation regions etc.

\begin{figure*}
\centering
\includegraphics[width=\textwidth]{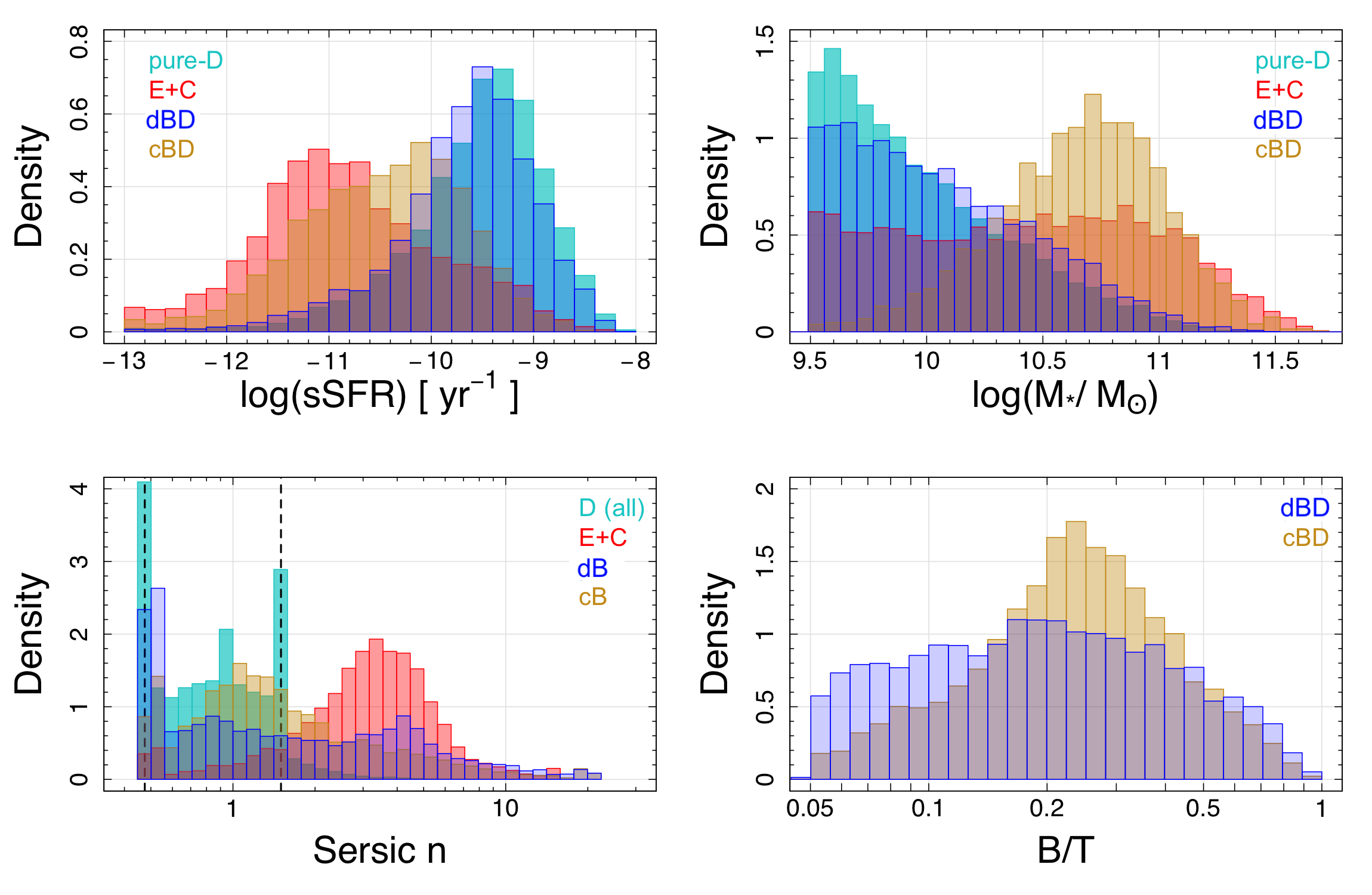}
\caption{PDF of the global sSFR, and total stellar mass (top panels), as well as component S\'ersic index (bottom left) and B/T of the double component systems (bottom right). Vertical dashed lines on the bottom left panel show our limits on the S\'ersic index of disk in double component fitting. A few bins seen beyond this buff=er represent our pure disk systems fitted by a single S\'ersic for which our buffer range is wider. See the text for more details. }
\label{fig:sSSFR_Mass}
\end{figure*}

\begin{figure*}
	\centering
	\includegraphics[width = 0.95\textwidth, angle = 0]{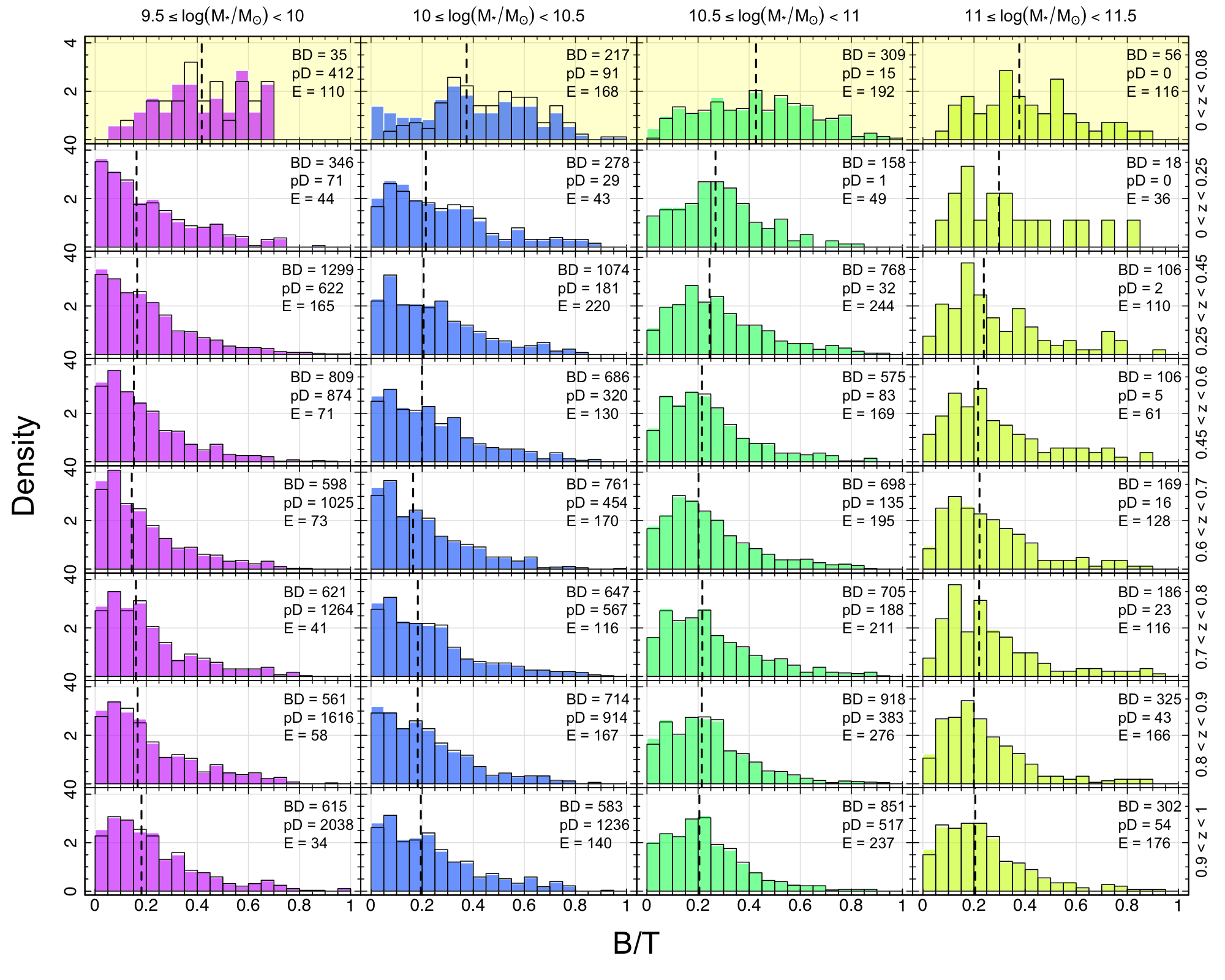}
	\caption{ The evolution of B/T as a function of total stellar mass (columns) and redshift (rows). The inset numbers indicate the number of pure disk (pD), double-component (BD) and elliptical (E) systems in each bin. The first row highlighted by yellow shows the histograms of B/T for $z=0$ GAMA galaxies. Dashed lines show the median values. Empty histograms with black borders represent systems with $\rm{R}_e > 0.25$ kpc while the background histograms show the total distribution in each bin.}
	\label{fig:BT}
\end{figure*}

\begin{figure*}
	\centering
	\includegraphics[width = \textwidth, angle = 0]{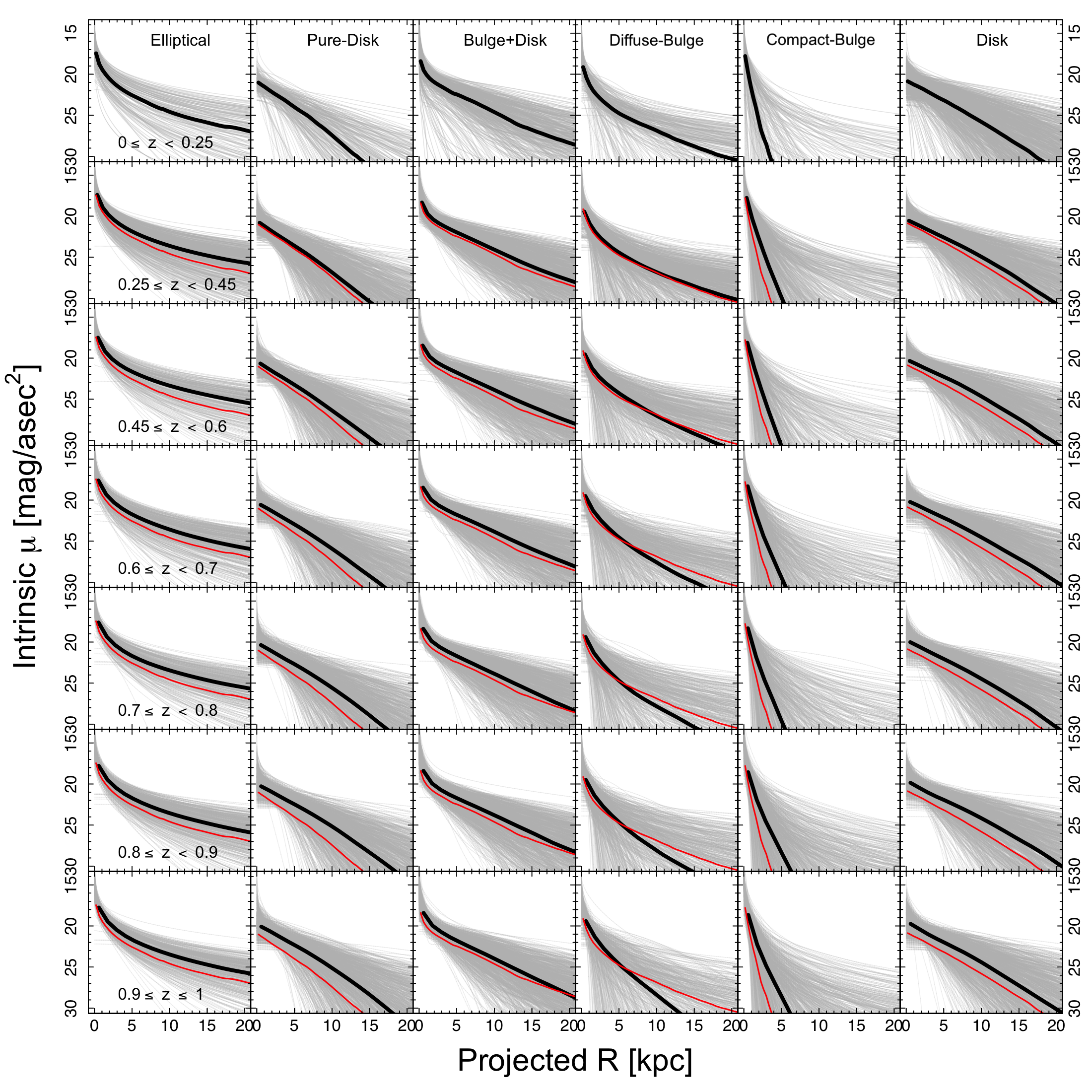}
	\caption{One-dimensional radial profile of different morphological types, as well as bulges and disks at different redshifts. Gray curves are 1000 profiles in each category (less in case of ellipticals). The overlaid thick profile is the median curve, 
	while thin red curves represent the median profiles at our lowest redshift range, i.e., $0.0 < z < 0.25$. This plot only includes the component mass of $> 10^{9.5} M_\odot$. }
	\label{fig:R_profiles}
\end{figure*}

\subsection{Discussion of our structural decompositions} 
\label{sec:str_dec_discussion}

Figure \ref{fig:B_D_scatterPlot} shows the relation between the physical half light radius, $\rm{R}_e$, of bulges and disks with redshift, components' stellar mass and I-mag. Note that we show the pixel size of HST/ACS, 0.03 arcsec/pixel (lower boundary in the left panel) to highlight that our measured structures, bulges in particular, are predominantly larger than the size of pixels, although we note a small number of unresolved bulges, at higher redshift, at very low stellar mass and at very faint apparent magnitudes. In this work, we will be limiting our studies to components with stellar masses $> 10^{9.5} M_{\odot}$ which removes most of the unresolved bulges. Given the 5$\sigma$ limiting depth of the COSMOS ACS F814W filter to be 27.2 (AB in a $0.''24$ diameter aperture, \citealt{Koekemoer07}) for point sources, the right panel of Figure \ref{fig:B_D_scatterPlot} indicates that all flux of our components are within the flux limit of the imaging data.

Figure \ref{fig:sSSFR_Mass} shows our final galaxy populations in various systemic observable or intrinsic parameter spaces. As expected, pure disk galaxies (D) are the least massive with the highest specific star-formation rates (sSFR). As one might also expect, elliptical galaxies (E) are the most massive with the lowest sSFR systems. This Figure also shows that disk galaxies containing a cB (cBD) do, in general, have lower sSFR, are more massive, and have higher B/T values than systems containing a dB (dBD). Figure \ref{fig:sSSFR_Mass} further indicates that ellipticals dominate the higher values of the systemic S\'ersic index (bottom left panel), $n_b \simeq 4$, indicating near de Vaucouleurs light profile (\citealt{deVaucouleurs48}), while disks occupy lower regions around $n_d \simeq 1$, indicating near exponential light profile. Interestingly, we do not find a significant discrimination between S\'ersic indices of dBs and cBs. In fact, we find cBs' S\'ersic index peaked around $n = 1$ and dBs' peaked at both $n = 1$ and 4.
Therefore, bulge S\'ersic indices, extend across the whole parameter space from 0.5 to 10, showing no clear correlation between the systemic S\'ersic index and the bulge morphology. We note that the systemic S\'ersic index does show some differentiation, but does not map well to bulge type, as noted earlier and reported in recent IFU studies (e.g., \citealt{Krajnovic13}; \citealt{Zhu18}; \citealt{Schulze18}; \citealt{Zhu20}; \citealt{Oh20}). We, however, do not rule out some uncertainties due to our dB/cB separation technique. 

Note that since we limit the range of S\'ersic indices of disks and bulges to $0.5 < n_d < 1.5$ and $0.5 < n_b < 20$, respectively, we find some fits trapped at lower or higher limits (see the bottom left panel of Figure \ref{fig:sSSFR_Mass}). One might decide to solve this by extending the buffer to give the mathematical modelling freedom to explore a wider space. Highlighting that not every mathematically-preferred optimised model is necessarily synonymous with the most physically valid ones, we decided to keep the parameters in a physically induced range following \cite{Cook19}. For example, one expects a stable disk to have a S\'ersic index close to unity. As a consequence of this buffer selection, we find $n_d$ histogram (cyan) also presenting two peaks on the boundaries ($n = 0.5$ and $1.5$, bottom left panel of Figure \ref{fig:sSSFR_Mass}). 

In Figure \ref{fig:BT}, we further inspect the correlation between B/T and stellar mass as well as redshift. We select our redshift binning extending from $z=0$ to $z=1$ similar to \cite{Hashemizadeh21}. The first row highlighted with yellow shows the B/T distribution of GAMA galaxies. The Figure shows that massive galaxies typically have more significant bulges, i.e., larger B/T. It also indicates that B/T is, however, stable throughout time. Most noticeable is the rise in lower B/T systems in the lowest mass bin, potentially this may be due to some classification bias with very small bulges at very low mass intervals at high-$z$ becoming harder to visually identify. However, we note the opposite trend in the most massive galaxies. The HST resolution is given by the black line on the left panel of Figure~\ref{fig:B_D_scatterPlot}, and in general very few bulges are at this limit, suggesting that the increase in low B/T systems at low-redshift may be genuine. However, we cannot fully rule out some other bias. Ultimately, according to our pixel size ($0.03$ arcsec), we are only able to resolve bulges with $R_e > 0.25$ kpc across all redshift intervals (see black line in Figure~\ref{fig:B_D_scatterPlot}). To explore whether this bias is significant, in Figure \ref{fig:BT} we also show the results if we impose a uniform $R_e > 0.25$ kpc limit as the black line histograms, and while we do see a modest change in the very low-B/T objects in the lowest mass and redshift bin, the change is modest, and hence we conclude that the growth in low mass bulges towards lower redshift is real.

Note that for the GAMA data in the lowest redshift bin we find a more extended B/T range with a larger median value of the B/T. We note that the GAMA decompositions are still under review and not yet published.

Finally, Figure \ref{fig:R_profiles} shows a random selection of 1D component profiles, with component masses above $10^{9.5}$M$_{\odot}$ and indicative of our science analysis sample. Note that pure-disk, here, refers to galaxies visually classified as a pure-disk morphology, while disk refers to the disk component of bulge+disk systems. We convert the apparent surface brightness to the intrinsic surface brightness (SB) by correcting for $(1+z)^4$ SB dimming. Thick black curves represent the median profile for each subset, and red curves show the redshift zero fit. Our initial impression, is that there appears to be a marginal contraction (fading) in almost all structures likely due to merging galaxies of all stellar masses here. 

\section{The evolution of the SMF since \lowercase{$z$} $= 1$} 
\label{Sec:SMF_evol}

\begin{landscape}
\begin{figure}
	\centering
	\includegraphics[width = 1.3\textwidth, angle = 0]{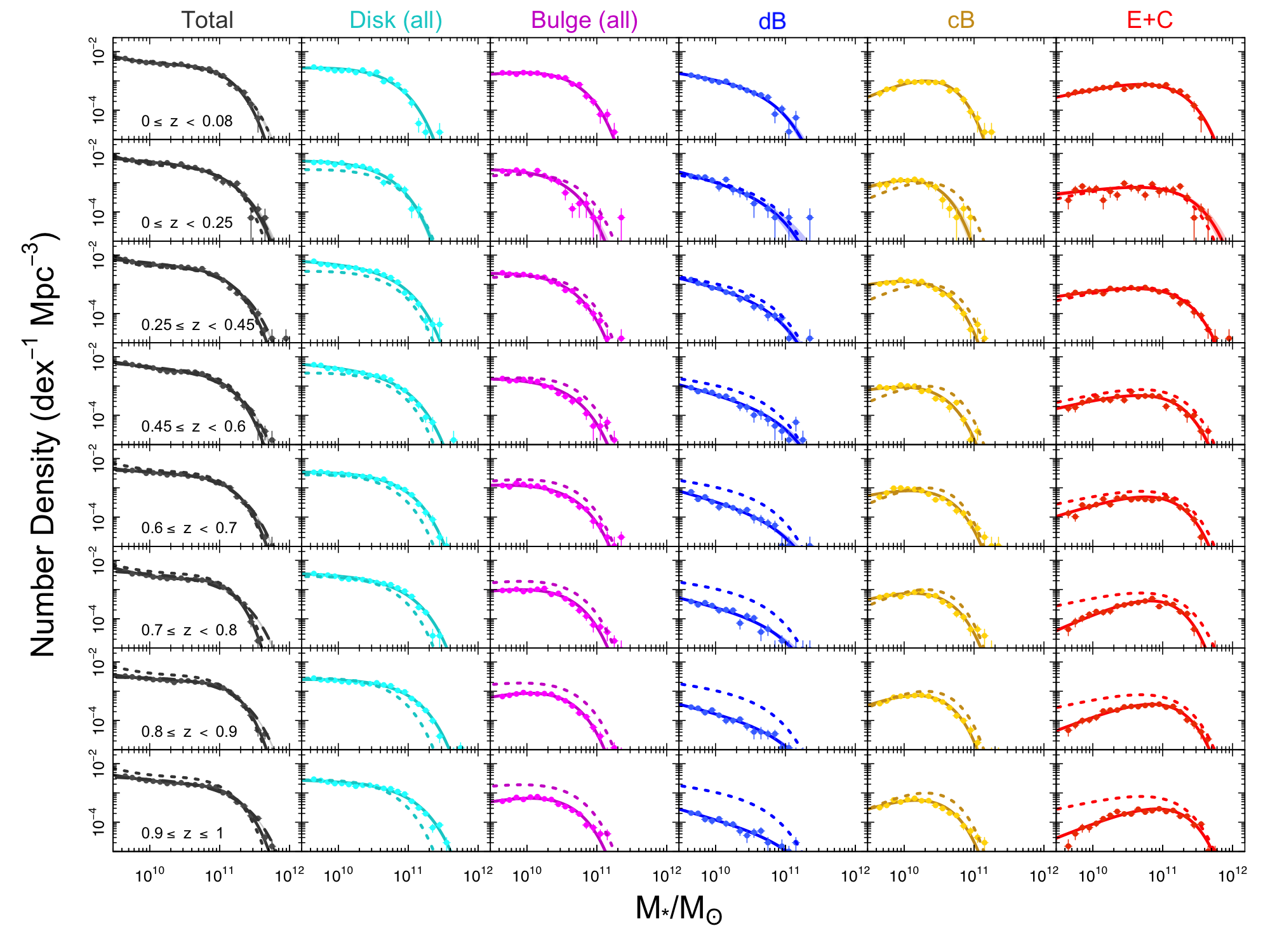}
	\caption{ The SMF of components in 8 redshift bins. The local GAMA SMFs are shown in the top row highlighted with yellow. Data points represent galaxy counts in each of equal-size stellar mass bins. Width of stellar mass bins are shown as horizontal bars on data points. Vertical bars show poisson errors. Shaded regions around the best fit curves are 68 per cent confidence regions. The over-plotted dotted curves represent the GAMA SMFs ($0 < z < 0.08$). Note that in the total SMFs (left column) solid lines represent our double Schechter fits to data that are very close to single Schechter fits (dashed curves).}
	\label{fig:Mfunc_Str}
\end{figure}
\end{landscape}

In \cite{Hashemizadeh21}, we showed that the volume-corrected distribution of morphologically subdivided stellar-mass for the D10/ACS sample is well described by single \cite{Schechter76} functions, as the mass range ($> 10^{9.5}M_{\odot}$) probed does not extend significantly beyond where a turn-up starts to be seen at around $10^{9.5}M_{\odot}$, while the global SMF is shown to fit well with double Schechter function (e.g., \citealt{Baldry08}; \citealt{Pozzetti10}; \citealt{Baldry12} and \citealt{Wright17}). In the present work, we therefore, use the same double Schechter function to fit our total SMF (solid black lines in Figure \ref{fig:Mfunc_Str}) although for completeness we also show our single Schechter fits as dashed black lines. Note that as can be seen in Figure \ref{fig:Mfunc_Str}, all components can be fitted with single Schechter functions at all redshifts.

To derive our stellar-mass functions (SMF), we make use of the \texttt{dftools} package implemented in R (see \citealt{Obreschkow18}). 
In all cases the fitted SMFs describe the data well, see Figure \ref{fig:Mfunc_Str}, which shows the evolution of both the total SMFs and the SMF broken into structural types of disks (all; including both pure disk systems and disk components), bulges (all), dB, cB and E+C (ellipticals+compacts). In Figure \ref{fig:Mfunc_Str}, each row represents a distinct redshift range extending from $z=0$ to 1, as indicated on the panel. As mentioned earlier, similar to \cite{Hashemizadeh21}, we select our redshift bins to be $z = 0.0,0.25,0.45,0.6,0.7,0.8,0.9,1.0$. For comparison, we also present the new local GAMA SMFs ($0.0 < z < 0.08$) in the top row of Figure \ref{fig:Mfunc_Str}. Note that we use our GAMA visual morphological classifications to inform our low-$z$ structural SMFs while the separation of dBs and cBs follows an identical procedure for both GAMA and DEVILS data as discussed in Section \ref{sec:dB_cB_dist}.

The total and elliptical SMFs are essentially identical to that shown in figure 12 in \cite{Hashemizadeh21}.
Our SMF values also include a correction for the large scale structure (LSS) along the COSMOS sight-line, i.e., under- and over-densities in the COSMOS field in different redshift bins, as described in section 4.2 in \cite{Hashemizadeh21}. In brief, we determine an LSS correction by forcing the total stellar-mass density (SMD) to match a smooth spline fit to the data of \cite{Driver18}. We then apply our LSS correction factors in each redshift interval to all SMD trends (all components) by multiplying by the scale factor.    

Figure \ref{fig:Mfunc_Str} highlights that the total SMF grows since $z = 1$ at both the low- and high-mass ends. We also see a similar increase with cosmic time in the low-mass end of the disk SMF, but a decrease in their intermediate- to high-mass end. Interestingly, the bulge component and ellipticals show stronger evolution with time with the dB's and cB's growing strongly and uniformly at all masses (internal secular processes and minor mergers?), and ellipticals predominantly at intermediate to lower-masses (major mergers?). Noticeable in the total data is the emergence of a bump and plateau in the mass function at lower redshifts. This has also been noted in \cite{Robotham14} and \cite{Wright18}. Physical interpretations will be discussed in Section \ref{sec:discussion}.
 
\begin{figure}
	\centering
	\includegraphics[width = 0.48\textwidth, angle = 0]{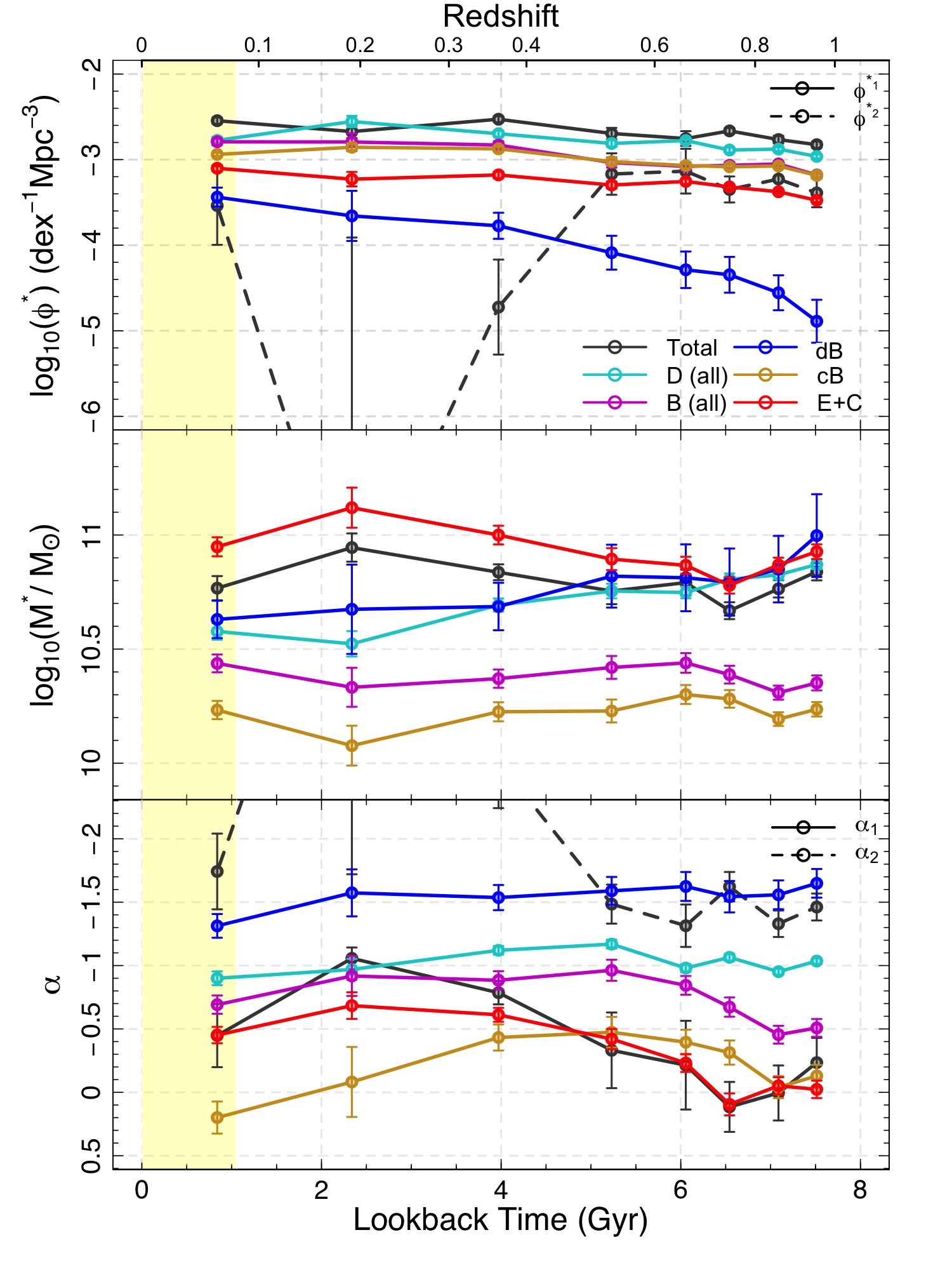}
	\caption{ The evolution of the best fit Schechter parameters from $z = 1$. Yellow band represents the time covered by our GAMA data. Dashed lines represent the parameters for our double Schechter fit to total and disk SMFs. For example, solid and dashed black lines in the top panel show the evolution of $\phi^{*}_{1}$ and $\phi^{*}_{2}$, respectively.  }
	\label{fig:Mfunc_par_evol}
\end{figure}

Finally, Figure \ref{fig:Mfunc_par_evol} shows the evolution of our best fit Schechter parameters as a function of lookback time for each component. The Schechter normalization parameter, $\phi^*$, of the total and disk population experiences a very slight increase since $z = 1$, while bulges' $\phi^*$ shows a small increase.
dBs occupy lower values and grow constantly over time while cBs and ellipticals experience a modest increase. Note that we also show the second parameters for our double Schechter functions (i.e., total and disk SMFs) as dashed lines. As expected, these are more fluctuated with larger errors.

The characteristic mass, $M^*$, also known as the knee or break mass of the total SMF is relatively stable while decreases for disks. The $M^*$ of all bulges and cBs evolves very little, while it decreases for dBs over the redshift range probed.

Lastly, $\alpha$, the faint end slope, is steepest for the dB population, but shows a modest decrease. The $\alpha$ of the total and disk population increases at all epochs except for the decline when transitioning to the GAMA data. It also increases for cB population by $z \sim 0.5$ and then declines to $z = 0$ while increasing for E+C types likely suggesting that low mass disks and bulges are growing/forming at all epochs. We report the tabulated version of our best Schechter function parameters in Table. \ref{tab:Str_MF_par}.

\begin{table*}
\centering
\caption{Best Schechter fit parameters of total and structural SMF in different redshift bins as well as the integrated SMD, i.e., $\mathrm{log}_{10}(\rho^*$) and the stellar baryon fraction, i.e., $f_s$. For completeness we report the $\rho^*$ values for integration over the stellar mass ranges of $0-\infty$ and $10^{9.5}-\infty$. $f_s$ is calculated using our main integration range ($0-\infty$).}
\begin{adjustbox}{scale = 0.62}
\begin{tabular}{lcccccccc}
\firsthline \firsthline \\
$z$-bins         &  $0.0 \leq z < 0.08$  &  $0.0 \leq z < 0.25$ &  $0.25 \leq z < 0.45$ &  $0.45 \leq z < 0.60$ &  $0.60 \leq z < 0.70$ &  $0.70 \leq z < 0.80$ &  $0.80 \leq z < 0.90$ &  $0.90 \leq z \leq 1.00$ \\ \\ \hline 
                   &   \multicolumn{8}{c}{\textbf{Total (Double Schechter)}}  \\
\cline{2-9} \\
$\mathrm{log}_{10}(\Phi^*_1)$      & $-2.55\pm0.04$  & $-2.67\pm0.08$   & $-2.53\pm0.03$ & $-2.69\pm0.07$ & $2.75\pm0.09$   & $-2.66\pm0.03$ & $-2.77\pm0.05$ & $-2.83\pm0.04$ \\ \\
$\mathrm{log}_{10}(M^*)$           & $10.77\pm0.05$  & $10.95\pm0.06$   & $10.84\pm0.03$ & $10.75\pm0.06$ & $10.79\pm0.06$  & $10.67\pm0.04$ & $10.76\pm0.04$ & $10.84\pm0.04$  \\ \\
$\alpha_1$                         & $-0.45\pm0.25$  & $-1.06\pm0.09$   & $-0.79\pm0.09$ & $-0.33\pm0.30$ & $-0.21\pm0.35$  & $0.12\pm0.20$  & $0.01\pm0.22$  & $-0.23\pm0.20$ \\ \\
$\mathrm{log}_{10}(\Phi^*_2)$      & $-3.54\pm0.46$  & $-8.40\pm4.48$   & $-4.72\pm0.55$ & $-3.17\pm0.24$ & $-3.13\pm0.26$  & $-3.35\pm0.15$ & $-3.23\pm0.15$ & $-3.39\pm0.17$   \\ \\
$\alpha_2$                         & $-1.74\pm0.30$  & $-4.77\pm3.05$   & $-2.61\pm0.37$ & $-1.48\pm0.15$ & $-1.31\pm0.17$  & $-1.62\pm0.12$ & $-1.33\pm0.10$ & $-1.46\pm0.11$  \\ \\ 
$\mathrm{log}_{10}(\rho^*) [10^{9.5}-\infty]$        & $8.245\pm0.08$   & $8.272\pm0.14$  & $8.235\pm0.09$ & $8.185\pm0.10$ & $8.183\pm0.13$ & $8.133\pm0.09$ & $8.143\pm0.09$ & $8.119\pm0.09$  \\ \\
$\mathrm{log}_{10}(\rho^*) [0-\infty]$               & $8.318\pm0.13$   & $8.288\pm0.24$  & $8.259\pm0.13$ & $8.231\pm0.23$ & $8.207\pm0.12$ & $8.192\pm0.11$ & $8.165\pm0.09$ & $8.151\pm0.09$  \\ \\
$f_s$                                                & $0.0349\pm0.0118$  & $0.0326\pm0.0239$  & $0.0305\pm0.0109$ & $0.0286\pm0.0201$ & $0.0271\pm0.0086$ & $0.0261\pm0.0077$ & $0.0246\pm0.0057$ & $0.0238\pm0.0055$ \\ \\ \hline
                    &   \multicolumn{8}{c}{\textbf{Total (Single Schechter)}}  \\
\cline{2-9} \\
$\mathrm{log}_{10}(\Phi^*)$         & $-2.74\pm0.03$   & $-2.74\pm0.06$   & $-2.74\pm0.03$ & $-2.81\pm0.03$ & $-2.79\pm0.03$ & $-2.91\pm0.03$ & $-2.86\pm0.02$ & $-2.98\pm0.2$ \\ \\
$\mathrm{log}_{10}(M^*)$            & $10.98\pm0.03$   & $10.99\pm0.05$   & $11.00\pm0.03$ & $10.97\pm0.03$ & $10.98\pm0.02$ & $11.03\pm0.02$ & $11.02\pm0.02$ & $11.09\pm0.02$ \\\\
$\alpha$                            & $-1.09\pm0.03$   & $-1.14\pm0.05$   & $-1.16\pm0.02$ & $-1.16\pm0.03$ & $-1.03\pm0.02$ & $-1.11\pm0.02$ & $-0.99\pm0.02$ & $-1.10\pm0.02$ \\ \\ 
$\mathrm{log}_{10}(\rho^*)[10^{9.5}-\infty]$          & $8.246\pm0.07$  & $8.272\pm0.11$   & $8.284\pm0.08$  & $8.185\pm0.08$  & $8.183\pm0.09$ & $8.133\pm0.09$  & $8.144\pm0.09$   & $8.120\pm0.09$ \\\\
$\mathrm{log}_{10}(\rho^*) [0-\infty]$                & $8.266\pm0.07$  & $8.296\pm0.11$  & $8.310\pm0.08$ & $8.212\pm0.08$ & $8.199\pm0.09$ & $8.153\pm0.09$ & $8.156\pm0.09$ & $8.137\pm0.09$  \\ \\
$f_s$                                                 & $0.0310\pm0.0061$  & $0.0332\pm0.0093$  & $0.0343\pm0.0064$ & $0.0274\pm0.0061$ & $0.0266\pm0.0059$ & $0.0239\pm0.0048$ & $0.0241\pm0.0052$ & $0.0230\pm0.0049$ \\ \\ \hline
                    &   \multicolumn{8}{c}{\textbf{Disk (all)}}  \\
 \cline{2-9} \\
$\mathrm{log}_{10}(\Phi^*)$         & $-2.96\pm0.02$  & $-2.55\pm0.07$   & $-2.70\pm0.04$ & $-2.81\pm0.04$ & $-2.78\pm0.03$ & $-2.89\pm0.03$ & $-2.88\pm0.02$ & $-2.96\pm0.02$ \\ \\
$\mathrm{log}_{10}(M^*)$            & $10.87\pm0.02$  & $10.52\pm0.06$   & $10.69\pm0.03$ & $10.75\pm0.03$ & $10.75\pm0.03$ & $10.81\pm0.03$ & $10.83\pm0.02$ & $10.87\pm0.02$ \\ \\
$\alpha$                            & $-1.04\pm0.02$  & $-0.97\pm0.08$   & $-1.12\pm0.03$ & $-1.17\pm0.04$ & $-0.98\pm0.03$ & $-1.06\pm0.03$ & $-0.95\pm0.02$ & $-1.04\pm0.02$ \\ \\
$\mathrm{log}_{10}(\rho^*)[10^{9.5}-\infty]$          & $7.754\pm0.07$   & $7.923\pm0.11$   & $7.990\pm0.08$ & $7.954\pm0.08$ & $7.943\pm0.09$ & $7.908\pm0.09$ & $7.920\pm0.09$ & $7.896\pm0.09$  \\ \\
$\mathrm{log}_{10}(\rho^*)[0-\infty]$                 & $7.781\pm0.07$   & $7.961\pm0.11$   & $8.032\pm0.08$ & $7.997\pm0.08$ & $7.966\pm0.09$ & $7.935\pm0.09$ & $7.937\pm0.09$ & $7.917\pm0.09$  \\ \\
$f_s$                                                 & $0.0101\pm0.0018$ & $0.0154\pm0.0043$  & $0.0181\pm0.0038$ & $0.0167\pm0.0036$ & $0.0155\pm0.0038$ & $0.0145\pm0.0034$ & $0.0145\pm0.0033$ & $0.0139\pm0.0031$ \\ \\ \hline
                    &   \multicolumn{8}{c}{\textbf{Bulge (all)}}  \\
 \cline{2-9} \\
$\mathrm{log}_{10}(\Phi^*)$         & $-3.18\pm0.03$  & $-2.79\pm0.10$   & $-2.83\pm0.05$ & $-3.03\pm0.06$ & $-3.08\pm0.05$ & $-3.07\pm0.04$ & $-3.05\pm0.03$ & $-3.18\pm0.03$ \\ \\
$\mathrm{log}_{10}(M^*)$            & $10.35\pm0.03$  & $10.33\pm0.09$   & $10.37\pm0.04$ & $10.42\pm0.05$ & $10.44\pm0.04$ & $10.39\pm0.04$ & $10.31\pm0.03$ & $10.35\pm0.03$ \\ \\
$\alpha$                            & $-0.51\pm0.07$  & $-0.92\pm0.16$   & $-0.88\pm0.07$ & $-0.96\pm0.08$ & $-0.84\pm0.07$ & $-0.67\pm0.08$ & $-0.45\pm0.07$ & $-0.51\pm0.07$ \\ \\
$\mathrm{log}_{10}(\rho^*)[10^{9.5}-\infty]$          & $7.577\pm0.07$   & $7.469\pm0.11$ & $7.473\pm0.08$ & $7.329\pm0.09$ & $7.297\pm0.10$ & $7.250\pm0.09$ & $7.190\pm0.09$ & $7.103\pm0.09$  \\ \\
$\mathrm{log}_{10}(\rho^*)[0-\infty]$                 & $7.599\pm0.07$   & $7.520\pm0.12$ & $7.516\pm0.08$ & $7.377\pm0.09$ & $7.329\pm0.10$ & $7.273\pm0.09$ & $7.207\pm0.09$ & $7.120\pm0.09$  \\ \\
$f_s$                                                 & $0.0067\pm0.0012$ & $0.0056\pm0.0017$  & $0.0055\pm0.0012$ & $0.0040\pm0.0009$ & $0.0036\pm0.0009$ & $0.0031\pm0.0008$ & $0.0027\pm0.0006$ & $0.0022\pm0.0005$ \\ \\ \hline
                    &   \multicolumn{8}{c}{\textbf{Diffuse-Bulge}}  \\
 \cline{2-9} \\
$\mathrm{log}_{10}(\Phi^*)$         & $-3.44\pm0.11$  & $-3.66\pm0.3$   & $-3.77\pm0.2$ & $-4.09\pm0.20$ & $-4.29\pm0.21$ & $-4.35\pm0.21$ & $-4.56\pm0.20$ & $-4.89\pm0.25$ \\ \\
$\mathrm{log}_{10}(M^*)$            & $10.63\pm0.08$  & $10.67\pm0.2$   & $10.69\pm0.1$ & $10.82\pm0.14$ & $10.81\pm0.15$ & $10.79\pm0.15$ & $10.85\pm0.15$ & $11.00\pm0.18$ \\ \\
$\alpha$                            & $-1.31\pm0.09$  & $-1.57\pm0.2$   & $-1.54\pm0.1$ & $-1.59\pm0.11$ & $-1.62\pm0.11$ & $-1.54\pm0.12$ & $-1.56\pm0.11$ & $-1.65\pm0.11$ \\ \\
$\mathrm{log}_{10}(\rho^*)[10^{9.5}-\infty]$          & $7.226\pm0.08$  & $7.148\pm0.13$   & $7.032\pm0.09$ & $6.899\pm0.10$ & $6.707\pm0.11$ & $6.590\pm0.11$ & $6.453\pm0.11$ & $6.338\pm0.11$     \\ \\
$\mathrm{log}_{10}(\rho^*)[0-\infty]$                 & $7.312\pm0.09$  & $7.335\pm0.31$   & $7.194\pm0.13$ & $7.066\pm0.15$ & $6.899\pm0.23$ & $6.735\pm0.18$ & $6.597\pm0.16$ & $6.513\pm0.22$     \\ \\
$f_s$                                                 & $0.0035\pm0.0008$ & $0.0036\pm0.0037$  & $0.0026\pm0.0009$ & $0.0020\pm0.0011$ & $0.0013\pm0.0007$ & $0.0009\pm0.0004$ & $0.0007\pm0.0005$ & $0.0006\pm0.0005$     \\ \\ \hline
                    &   \multicolumn{8}{c}{\textbf{Compact-Bulge}}  \\
 \cline{2-9} \\
$\mathrm{log}_{10}(\Phi^*)$         & $-2.94\pm0.02$  & $-2.86\pm0.05$   & $-2.87\pm0.03$ & $-3.02\pm0.04$ & $-3.07\pm0.03$  & $-3.08\pm0.03$ & $-3.07\pm0.02$ & $-3.18\pm0.02$ \\ \\
$\mathrm{log}_{10}(M^*)$            & $10.23\pm0.04$  & $10.08\pm0.09$   & $10.23\pm0.04$ & $10.23\pm0.05$ & $10.30\pm0.04$  & $10.28\pm0.04$ & $10.19\pm0.03$ & $10.24\pm0.03$ \\ \\
$\alpha$                            & $0.20\pm0.13$   & $-0.08\pm0.28$   & $-0.43\pm0.10$ & $-0.47\pm0.12$ & $-0.40\pm0.10$  & $-0.31\pm0.10$  & $-0.04\pm0.09$  & $-0.13\pm0.09$ \\ \\
$\mathrm{log}_{10}(\rho^*)[10^{9.5}-\infty]$          & $7.332\pm0.08$   & $7.192\pm0.12$ & $7.280\pm0.09$ & $7.131\pm0.09$ & $7.168\pm0.10$ & $7.143\pm0.10$ & $7.103\pm0.09$ & $7.021\pm0.09$     \\ \\
$\mathrm{log}_{10}(\rho^*)[0-\infty]$                 & $7.336\pm0.08$   & $7.208\pm0.13$ & $7.301\pm0.09$ & $7.154\pm0.09$ & $7.183\pm0.10$ & $7.156\pm0.10$ & $7.111\pm0.09$ & $7.030\pm0.09$     \\ \\
$f_s$                                                 & $0.0036\pm0.0007$ & $0.0027\pm0.0009$  & $0.0034\pm0.0008$ & $0.0024\pm0.0006$ & $0.0026\pm0.0007$ & $0.0024\pm0.0006$ & $0.0022\pm0.0005$ & $0.0018\pm0.0004$     \\ \\ \hline
                   &   \multicolumn{8}{c}{\textbf{Elliptical + Compact}}  \\
 \cline{2-9} \\
$\mathrm{log}_{10}(\Phi^*)$         & $-3.10\pm0.04$  & $-3.23\pm0.08$   & $-3.18\pm0.04$ & $-3.30\pm0.04$ & $-3.25\pm0.03$ & $-3.32\pm0.02$ & $-3.38\pm0.02$ & $-3.48\pm0.02$ \\ \\
$\mathrm{log}_{10}(M^*)$            & $10.95\pm0.04$  & $11.12\pm0.09$   & $11.00\pm0.04$ & $10.89\pm0.05$ & $10.87\pm0.04$ & $10.78\pm0.04$ & $10.87\pm0.03$ & $10.93\pm0.03$ \\ \\
$\alpha$                            & $-0.45\pm0.06$  & $-0.68\pm0.10$   & $-0.61\pm0.05$ & $-0.42\pm0.08$ & $-0.23\pm0.07$ & $0.10\pm0.09$ & $-0.05\pm0.07$ & $-0.02\pm0.07$ \\ \\
$\mathrm{log}_{10}(\rho^*)[10^{9.5}-\infty]$          & $7.795\pm0.08$  & $7.840\pm0.14$   & $7.766\pm0.09$ & $7.546\pm0.10$ & $7.577\pm0.10$ & $7.478\pm0.10$ & $7.484\pm0.10$ & $7.447\pm0.10$ \\ \\
$\mathrm{log}_{10}(\rho^*)[0-\infty]$                 & $7.796\pm0.08$  & $7.843\pm0.14$   & $7.769\pm0.09$ & $7.548\pm0.10$ & $7.578\pm0.10$ & $7.479\pm0.10$ & $7.484\pm0.10$ & $7.448\pm0.10$ \\ \\
$f_s$                                                 & $0.0105\pm0.0022$ & $0.0117\pm0.0042$  & $0.0099\pm0.0023$ & $0.0059\pm0.0015$ & $0.0064\pm0.0017$ & $0.0051\pm0.0013$ & $0.0051\pm0.0013$ & $0.0047\pm0.0012$     \\ \\ \hline

\lasthline
\end{tabular}
\end{adjustbox}
\label{tab:Str_MF_par}
\end{table*}

\section{The Evolution of the SMD since \lowercase{$z$} $= 1$} 
\label{sec:rho}

Having derived the SMF of different galaxy components, we can now determine the stellar mass density (SMD) distribution for each population and finally calculate the total integrated stellar mass locked in each component. 
To construct the total SMD, $\rho^*$, for each type, we integrate under the distribution of the SMDs over all stellar masses from 0 to $\infty$. We note that our measurements of $\rho^*$ are generally robust to the integration range and in Table \ref{tab:Str_MF_par} we report both the $\rho^*$ derived from integrating from $10^{9.5}$ to infinity and when extrapolating to zero mass. In almost all cases, the extrapolated portion contains less than 5\% of the measured SMD, except in the case of dBs where it rises to 33\% at higher redshifts. This supports our assertion that the majority of mass is captured by the galaxies studied in our analysis.

\begin{figure*}
	\centering
	\includegraphics[width=\textwidth]{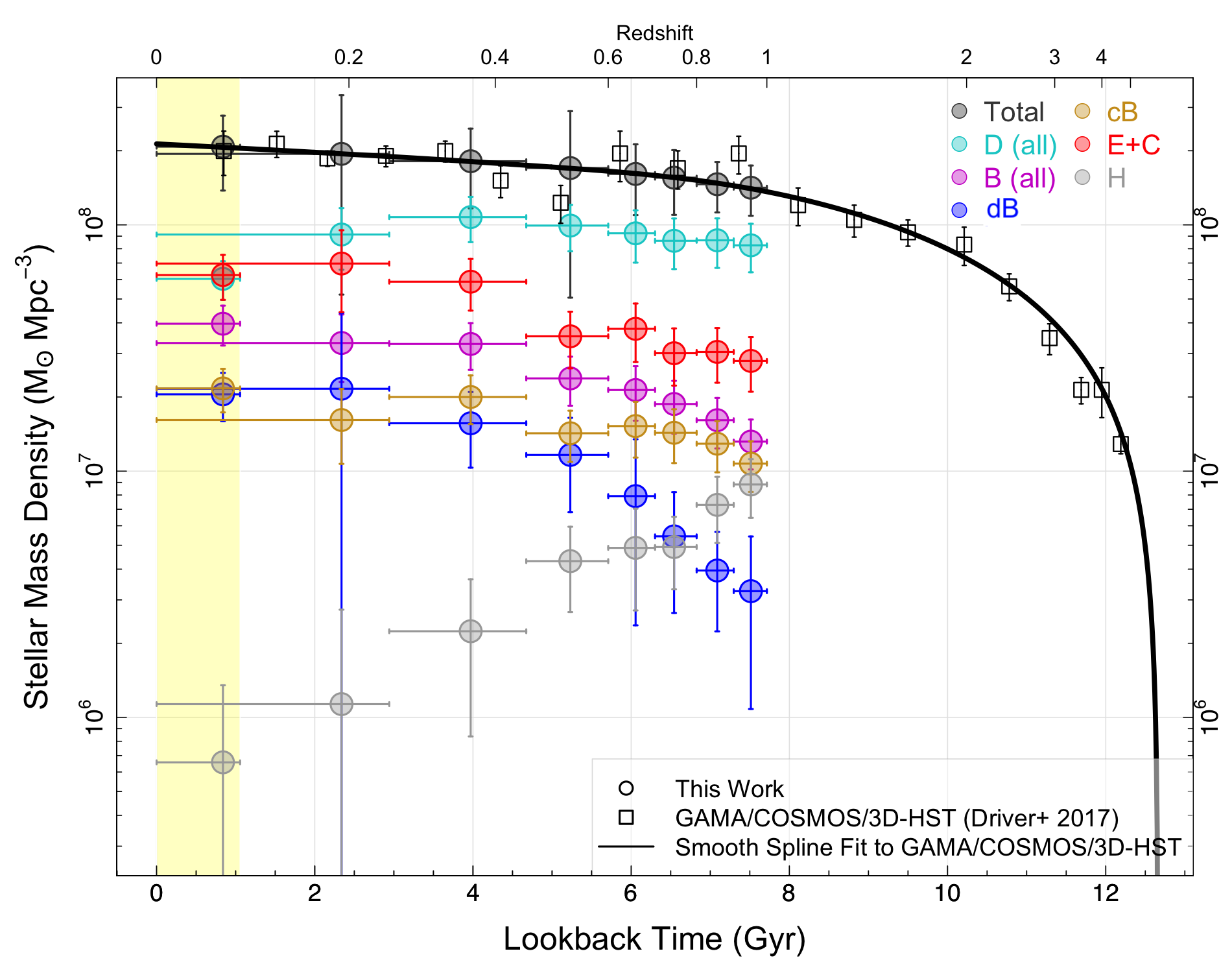}
    \caption{The evolution of the total and structural SMD, $\rho_{*}$, in the last 8 Gyr of the cosmic age. Colour codes are similar to previous plots. Vertical bars on the points show all errors including, fits and Poisson errors together with the classification and the cosmic variance error within the associated redshift bins taken from \protect\cite{Driver10}. Horizontal bars show the redshift ranges, while the data points are plotted at the mean redshift. The correction for the cosmic large scale structure is applied to the SMD in each redshift, as discussed in text and \protect\cite{Hashemizadeh21}. The local SMDs from GAMA are highlighted with yellow band. Note that we combine Cs with Es (E+C) here. See the text for more details. }
	\label{fig:MassBuildUp}
\end{figure*}

Figure \ref{fig:MassBuildUp} shows the evolution of the extrapolated $\rho_*$ values for each component. As a reminder, the SMDs include the correction for the large scale structure as described in \cite{Hashemizadeh21}. We also show the measurement from local galaxies as obtained from our GAMA measurements using identical methods, classifiers and techniques highlighted by a yellow transparent band (see Hashemizadeh et al. 2021 - PhD thesis). The error bars include all sources of uncertainties including classification, Poisson and fitting errors as well as the Cosmic Variance (CV) obtained as described in \cite{Hashemizadeh21}. 

\begin{figure}
	\centering
	\includegraphics[width = 0.5\textwidth, angle = 0]{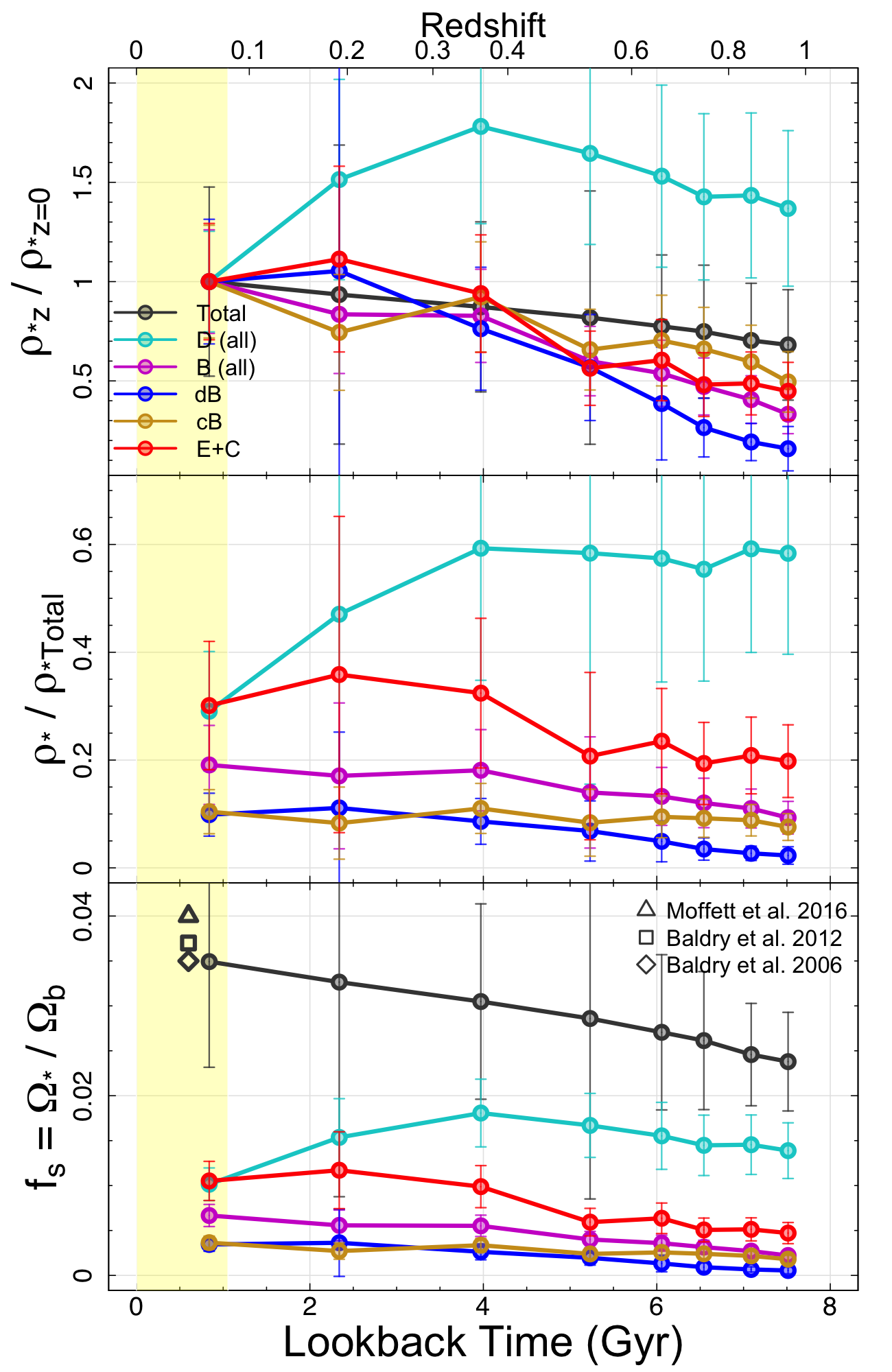}
	\caption{Top panel: variation of the stellar mass density, $\rho_{*}$, indicating the fraction of final stellar mass density (at $\overline{z} = 0$) assembled or lost by each redshift, i.e., $\rho_{*z}/\rho_{*z=0}$. Middle panel: shows the ratio of the structural SMDs to the total SMD at each redshift, i.e., $\rho_{*}/\rho_{*Total}$.
	Bottom panel: shows the evolution of the baryon fraction in form of stars in each galaxy component, i.e., $f_s = \Omega_*/\Omega_b$. Yellow band represents the time covered by our GAMA data.}
	\label{fig:fs}
\end{figure}

By splitting the total SMD into separate contributions of bulges and disks, we find that the disk component dominates the SMD of the Universe at all epochs except $z=0$ ($\sim 50\%$ of the total SMD, on average; see middle panel of Figure \ref{fig:fs}). The SMD of disks increases by a factor of $\sim 1.3$ over the interval $z = 1.0$ to $z \sim 0.35$, then declines by a factor of $\sim 0.57$ to $z = 0$. The SMD of (all) bulges experiences more significant growth at all epochs increasing by a factor of $\sim 3$ from $z = 1$ to 0. 

Subdividing our bulges into dB and cB, we find that cBs dominate over dBs in terms of the mass density, particularly at $z > 0.35$. 
However, we note that the dB SMD grows dramatically ($\sim \times 6.3$) from $z = 1$ to 0, while cBs grow consistently by a factor of $\sim 1.86$ to $z \sim 0.35$ and thereafter flatten.
If dBs can be considered redistributed disk material then there is some justification to fold the dB SMD measurements into the disk category. In this case, the disk+dB contribution at $z=0$ constitutes $40\%$ of the total SMD.
Finally, we note that the elliptical population also grows by a factor of $\sim \times 2.23$ over this time interval. 

To summarise these findings, Figure \ref{fig:fs} shows the evolution of the SMD as fractional changes (i.e., $z \sim 0$ ($\rho_{*z}/\rho_{*z=0}$) for the different structures (top panel), together with the ratio of the SMD of each component to the total (middle panel). The bottom panel shows the evolution of the stellar baryon fraction, i.e., $f_s = \Omega_*/\Omega_b$ where $\Omega_* = \rho_*/\rho_c$ and $\Omega_b = 0.0493$ \citep{Planck20} with the critical density of the Universe assumed to be $\rho_c = 1.21 \times 10^{11} \mathrm{M}_{\odot} \mathrm{Mpc}^{-3}$ at our GAMA median redshift ($\overline{z} \sim 0.06$) in a 737 cosmology. We report all the $f_s$ values at higher redshifts in Table \ref{tab:Str_MF_par}.

We find that, unsurprisingly, all galaxy components except disk grow in stellar mass density from $z = 1$ to $0$. dBs show the largest fractional mass growth, but overall dBs still contribute the least to the total SMD ($\sim 2-11$\% at z = 1-0; see Figure~\ref{fig:fs}). The middle and lower panels of Figure \ref{fig:fs} indicate that disks have the largest contribution to the total SMD and the stellar baryon fraction at all epochs, but interestingly, they have decreased their contribution to the total SMD over the last 4 billion years. This declining contribution is mirrored as an increased significance in the elliptical and compact bulge populations that show comparable growth. At face value it appears that $z = 1-0.36$ represents the phase where disk population grows more slower than bulges and a period of both secular evolution (growing diffuse-bulges) and the growth and/or emergence of both ellipticals and compact bulges (see their contributions to the total SMD in the middle panel of Figure \ref{fig:fs}).

\section{Discussion}
\label{sec:discussion}

Section \ref{sec:rho} reports the factual measurements of our structural stellar mass densities. Here we try to provide some interpretation of these measurements in the context of galaxy formation. However, before commencing, it is worth highlighting the various caveats in play. 

\begin{enumerate}
\item
We have identified that the HST PSF is less stable than one would like, which will introduce errors in bulge shape measurements and in particular the recovered S\'ersic indices. 

\item
Bulge-disk decomposition is also fraught with concerns over the minimisation algorithm becoming trapped in a local minimum or guided by the initial conditions. Our new Bayesian MCMC code (ProFit), is specifically designed to overcome this and Figures~\ref{fig:Conv_test_HST} \& \ref{fig:Conv_test_GAMA} suggest this is the case. 

\item
Whether a galaxy requires decomposition into two-components (or one-component is appropriate to capture the true radial profile) is reliant on our eyeball classifications from \cite{Hashemizadeh21}. While these demonstrate greater than 90 per cent consistency across our three classifiers for all redshifts and masses, we cannot rule out systematic biases with redshift that are influencing all our classifiers in the same way. Certainly the smoothness of the data suggests random errors are not dominating and the consistency of the classifications, while not ruling out some bias, would suggest it is secondary and likely modifying but not driving the trends seen. 

\item
We accept that our bulge measurements are likely measuring ``bulge complexes'' and our classification process is most likely sifting the bulges into diffuse or compact based on the dominant component. It is likely that many of our bulges contain multiple components, e.g., bars, peanut/boxy bulges, nuclear disks, nuclei, and compact bulges. In due course it may be possible, with higher resolution higher signal-to-noise or IFU data to disentangle further, however here we believe that taking the simple approach of classifying the dominant component will introduce less uncertainty than trying to fit multiple components given the variable PSF and that we are working at the resolution limit. 

\item
No attempt has been made to correct for the influence of dust attenuation which we know is more severe at higher redshifts due to elevated star-formation and at shorter wavelengths because of stronger attenuation. In due course, with JWST mid-IR observations of selected galaxies, this issue could be explored. 
\item
We note that our stellar mass measurements for bulges and disks are necessarily based on applying a simple B/T multiplier due to having only a single HST band. In reality, some bulges and disks will have a range of mass-to-light ratios and we expect that this will introduce significant errors in individual galaxies and a modest bias in our aggregated masses. 

\end{enumerate}

The above caveats could combine or cancel in ways that may have a significant impact on our measured values but we don't believe that they are likely to drive the general trends we see in Figures \ref{fig:MassBuildUp} \& \ref{fig:fs}. One aspect that gives us confidence that this might be the case is the relatively smooth transition from the redshift trends seen in the DEVILS data to the GAMA data and in general we see that the GAMA data (shown in the yellow band) is consistent with extrapolations of the redshift trends seen in the DEVILS data. While these studies have used identical methodologies and codes the data quality is dramatically different (i.e., $1''$ vs $0.03''$ resolution), and strong biases dependent on the data quality would likely lead to discontinuities. All of these caveats provide rich potential for further study in future years with facilities such as JWST, ESO MUSE and KMOS. Moving forward we acknowledge these caveats but for the remainder of this section, assume that they are not driving the trends that we see. 

As our backdrop we are aware that the cosmic star-formation rate and galaxy major merger rates are both in significant decline by $\sim \times 6$ and $\sim \times 3$ since $z=1$ respectively (see \citealt{Driver18}, and \citealt{Robotham14}). Various studies (e.g., \citealt{vanderWel14}; \citealt{Trujillo11}; etc) have also reported a significant growth in galaxy sizes with decreasing redshift, and a relatively smooth and modest evolution in the overall stellar mass function (e.g., \citealt{Wright18}).

To now add to this overall picture, we find from Figure \ref{fig:MassBuildUp} a rise and fall in the total disk mass (peaking at $z=0.35$) and a consistent rise in the SMDs of the Elliptical and bulge populations. We note the Hard and Compact populations, which together contains relatively minimal mass is likely reflecting an increasing merger rate with lookback time (for Hards), as well as difficulties in the classification of low-mass systems and limiting resolution. However, given the mass involved we can for the moment ignore these classes. It is also notable that the rise of the Ellipticals, compacts and compact bulges is fairly similar, with the rise in the dB class the strongest, especially at high-$z$ and flattening at low-$z$. 

\subsection{How the integrated stellar mass distribution is shifting with redshift}
\begin{figure*}
	\centering
	\includegraphics[width = 1.5\columnwidth, angle = 0]{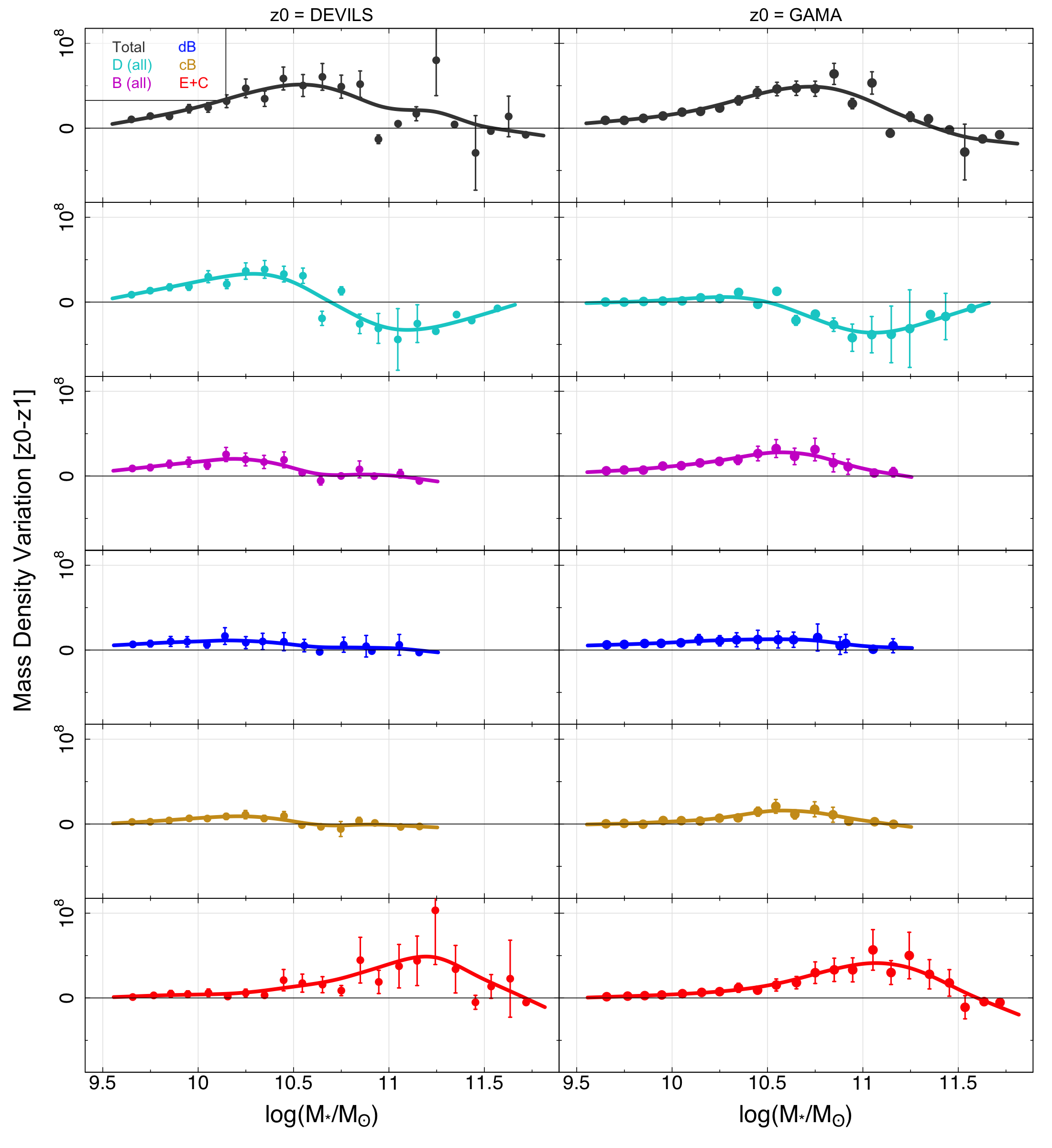}
	\caption{The variation of the SMD distribution (SMD$[z=0]-$SMD$[z=1]$, linear scale) as a function of stellar mass. Left column shows the variation when we take DEVILS lowest redshift bin ($0.0 \le z < 0.25$), while right panel indicates the variation when we take GAMA data as the lowest redshift bin ($0.0 < z \le 0.08$). Solid lines are smooth splines with degree of freedom of 6 fitted to the data points. }
	\label{fig:SMD_var}
\end{figure*}

Up until now we have focussed on the number evolution per mass interval and we have seen how the more dramatic changes are occurring at lower masses. However, while many galaxies may be involved, the total amount of mass can be surprisingly small. Hence here we now recast our results but with the focus more on how the total integrated mass within each mass interval has evolved from high to low redshift.  We show this on Figure ~\ref{fig:SMD_var} by plotting the {\it linear mass difference} in the SMD distributions in the highest and lowest redshift bins. This analysis very much highlights how mass, rather than galaxy numbers, have shifted. Not surprisingly small differences at high-mass are now highlighted as significant mass is involved, whereas large number differences at low-mass are supressed as the combined mass can be modest. The left-side panels of Figure~\ref{fig:SMD_var}, show the change relative to the lowest-$z$ DEVILS data (but where statistics are poor but data and methodologies identical), while the right-side panels use GAMA as the redshift zero reference data (where statistics are good and methodology similar, but data resolution is much coarser, 0.7$''$ versus 0.1$''$). In general, the left and right-side panels show a very similar picture which is reassuring. 

From Figure~\ref{fig:SMD_var} (top panel), we see that the dominant mass growth occurs around the knee of the mass-function and is fairly broad (i.e., $10^{10}-10^{11}$M$_{\odot}$) and this is consistent with the findings of \cite{Wright18} and \cite{Robotham14}. Now looking by components we see that the most interesting behaviour in the disks, where mass is lost at the high-mass end and gained at the lower mass-end. This loss at the high mass-end appears to be mirrored by the gain in mass at the Elliptical high-mass-end. This would seem to provide quite compelling evidence for Elliptical growth through the merger of high-mass disk systems, i.e. these disk mergers form Ellipticals hence and in doing so mass from one class to the other. In general, bulge growth while significant in Figure~\ref{fig:MassBuildUp} is less so in absolute mass terms (reflecting the log scaling in Figure~\ref{fig:MassBuildUp}). Nevertheless we can see that the growth in both the compact and diffuse-bulges is skewed slightly towards the lower mass end indicative of more subtle low-mass secular processes including accretion and redistribution.

\section{Summary and Conclusions} 
\label{sec:summary}

In this work, we have performed a 2D photometric structural decomposition of D10/ACS galaxies using high-resolution HST imaging. The goal is to provide a catalogue of credible structural measurements for galaxies at $z \leq 1$ and with $\mathrm{log}(M_*/M_\odot) \geq 9.5$ as defined in \cite{Hashemizadeh21}. In these two papers, we therefore provide the complex morphological and structural breakdown of the COSMOS galaxies. Catalogues are available within the DEVILS database, and will be released as part of the periodic DEVILS data releases. Here we summarize our results as follows:

\begin{description}

\item[$\bullet$] By performing several tests on the {\sc Tiny Tim} PSFs for HST/ACS we find that PSF uncertainties remain at the 10-20 per cent level, particularly within the central pixel, and attribute this to a combination of HST's periodic ``breathing'' (thermal expansion/contraction) and the difficulty in identifying the object centroid to very high precision. 

\item{$\bullet$} We use our decomposition pipeline, {\sc GRAFit} and the structural analysis code {\sc ProFit} (\citealt{Robotham17}) to fit three profile types to each of $\sim 35,000$ galaxy in the D10/ACS sample: a single S\'ersic, a S\'ersic+S\'ersic, and a S\'ersic+exponential disk. While we are unable to find a clear automated process for selecting the optimal profile, we find sensible outcomes when using our previous visual classifications to determine whether we should adopt a 1 or 2 component profiles. Where possible, we adopt double S\'ersic for the two component case, unless the bulge size exceeds the disk size, where we revert to a S\'erisc+exponential and confirm via visual inspection the veracity of the fit. 

\item{$\bullet$} The {\sc ProFit} Bayesian code together with the MCMC optimizations are demonstrated to be robust to initial conditions, and to provide good fits to more than 95 per cent of the sample, with poor fits generally arising when we have highly visually distorted or double-cored objects. This fraction of difficult to fit cases increases with look-back time.

\item{$\bullet$} We find that bulges generally fall into two categories, with a notable proportion forming a tight sequence in the stellar mass-half light radius ($M_*- \rm{R}_e$)-plane, and the remaining bulge systems scattered broadly across this plane. We attribute the compact distribution to be compact bulges, and the more dispersed distribution to be bulge-complexes which we refer to as diffuse-bulges (literally ``bulge-like'' or ``inner-disk'') and our selection is consistent with previous SDSS and GAMA results.

\item{$\bullet$} We report the B/T distributions in mass and redshift intervals, and find relatively strong trends in both directions, with what appears to be a strong emergence of low-B/T components in low mass systems at low redshifts. This appears to be robust to the evolution in our physical size limit at high-$z$ (i.e., $B_{\rm{R}_e} \sim 0.25$ kpc).

\item[$\bullet$] Moving from high- to low-redshift, the evolution of the stellar mass function for galaxy components (Figure \ref{fig:Mfunc_Str}) reveals an enhancement in the low-mass end while a modest growth in the high-mass end of the SMF as reported in \cite{Hashemizadeh21}. Subdividing by structural component we find:
\begin{description}
\item{(i)} Disk components increase their SMF number density at low-mass end, while showing a decrease at intermediate- and high-mass regions.
\item{(ii)} dBs and cBs experience significant growth at all mass intervals of their SMF.
\item{(iii)} Ellipticals show strong growth in the intermediate- and low-mass end of their SMF, and minimal evolution in their high mass end.
\end{description}

\item[$\bullet$] We report the distribution of stellar mass density and its growth from $z=1$ to $z=0$ and find:
\begin{description}
\item{(i)} Disks dominate the stellar mass density at all redshifts. 
The population increases their mass in  $z = 1-0.35$ and decreases gradually since then, their contribution to the total SMD declines from $\sim 60$ per cent to $\sim 32$ per cent over the whole redshift range. This appears to suggest an end to the epoch of disk growth.
\item{(ii)} The dB population grows in stellar mass by a factor of $\sim 6.3$ from $z = 1$ to $z = 0.0$. 
\item{(iii)} The cB population grows in stellar mass by a factor of $\sim 1.86$ from $z = 1$ to $z = 0.35$ and flattens its growth rate since then.
\item{(iv)} The Es contribute the most in the total stellar mass after disk population, growing in stellar mass by a factor of $\sim 2.23$ from $z = 1$ to $z = 0$.
\end{description}
\end{description}

We conclude that by performing a robust structural decomposition of D10/ACS galaxies using high-resolution HST imaging data we appear to unveil a Universe in which disk formation and growth has completed ($z = 1-0.35$), and then stalled/stabilised ($ z < 0.35$). The latter Universe is dominated by secular processes building diffuse-bulges, and ongoing minor and probably major mergers, consolidating and building mass in spheroidal structures (E's and cB's). The exact role of minor and major mergers is still unclear and somewhat hard to constrain in this study but a key goal of the DEVILS program. In addition, a critical factor we are unable to address here is the identification of the precise mechanisms or the role that may be played by dust attenuation. A key question is whether compact bulges are growing, or emerging as dust is dissipated. Future studies will explore the evolving dust content. Perhaps more critical is the need to directly measure the minor and major merger rates, which requires completion of the DEVILS redshift measurements, as well as understanding the neutral gas supply and accretion which will be unveiled through joint DEVILS MIGHTEE/LADUMA analysis. 
        
\section*{Acknowledgements}
DEVILS is an Australian project based around a spectroscopic campaign using the Anglo-Australian Telescope. The DEVILS input catalogue is generated from data taken as part of the ESO VISTA-VIDEO \citep{Jarvis13} and UltraVISTA \citep{McCracken12} surveys. DEVILS is part funded via Discovery Programs by the Australian Research Council and the participating institutions. The DEVILS website is \href{https://devilsurvey.org}{https://devilsurvey.org}. The DEVILS data is hosted and provided by AAO Data Central (\href{https://datacentral.org.au}{https://datacentral.org.au}). This work was supported by resources provided by The Pawsey Supercomputing Centre with funding from the Australian Government and the Government of Western Australia. We also gratefully acknowledge the contribution of the entire COSMOS collaboration consisting of more than 200 scientists. The HST COSMOS Treasury program was supported through NASA grant HST-GO-09822. SB and SPD acknowledge support by the Australian Research Council's funding scheme DP180103740. MS has been supported by the European Union's  Horizon 2020 research and innovation programme under the Maria Skłodowska-Curie (grant agreement No 754510), the National Science Centre of Poland (grant UMO-2016/23/N/ST9/02963) and by the Spanish Ministry of Science and Innovation through Juan de la Cierva-formacion program (reference FJC2018-038792-I). ASGR and LJMD acknowledge support from the Australian Research Council's Future Fellowship scheme (FT200100375 and FT200100055, respectively).

This work was made possible by the free and open R software (\citealt{R-Core-Team}).
A number of figures in this paper were generated using the R \texttt{magicaxis} package (\citealt{Robotham16b}). This work also makes use of the \texttt{celestial} package (\citealt{Robotham16a}) and \texttt{dftools} (\citealt{Obreschkow18}).

\subsection{Data Availability} 
\label{subsec:data_access}

The catalogues used in this paper are \texttt{D10VisualMoprhologyCat}, described in \cite{Hashemizadeh21}, and \texttt{DEVILS\_BD\_Decomp} and are held on the DEVILS database managed by AAO Data Central (\href{https://datacentral.org.au}{https://datacentral.org.au}). All imaging data are in the public domain and were downloaded from the the NASA/IPAC Infrared Science Archive (IRSA) web-page: \href{https://irsa.ipac.caltech.edu/data/COSMOS/images/acs\_2.0/I/}{irsa.ipac.caltech.edu/data/COSMOS/images/acs\_2.0/I/}. The main tools used in this study are {\sc ProFit} \citep{Robotham17} version 1.3.3 (available at: \href{https://github.com/ICRAR/ProFit}{https://github.com/ICRAR/ProFit}) and ProFound \citep{Robotham18} version 1.3.4 (available at: \href{https://github.com/asgr/ProFound}{https://github.com/asgr/ProFound}). We used {\sc Tiny Tim} version 6.3 to generate the HST/ACS Point Spread Function (PSF). We further use {\sc LaplacesDemon} version 1.3.4 implemented in {\sc R} available at: \href{https://github.com/LaplacesDemonR/LaplacesDemon}{https://github.com/LaplacesDemonR/LaplacesDemon}. Our structural decomposition pipeline, {\sc GRAFit}, is available at: \href{https://github.com/HoseinHashemi/GRAFit}{https://github.com/HoseinHashemi/GRAFit}.

\bibliographystyle{mnras}
\bibliography{library}

\appendix

\section{More details on the {\sc GRAFit} pipeline} 
\label{app:GRAFit_more}
In addition to Section \ref{sec:GRAFit}, here we extend on a few important measurements and prerequisites that {\sc GRAFit} takes into account.
    
\subsection{Dynamic cutout and sky estimation} 
\label{subsec:Sky est} 

\begin{figure} 
	\centering
	\includegraphics[width=\columnwidth]{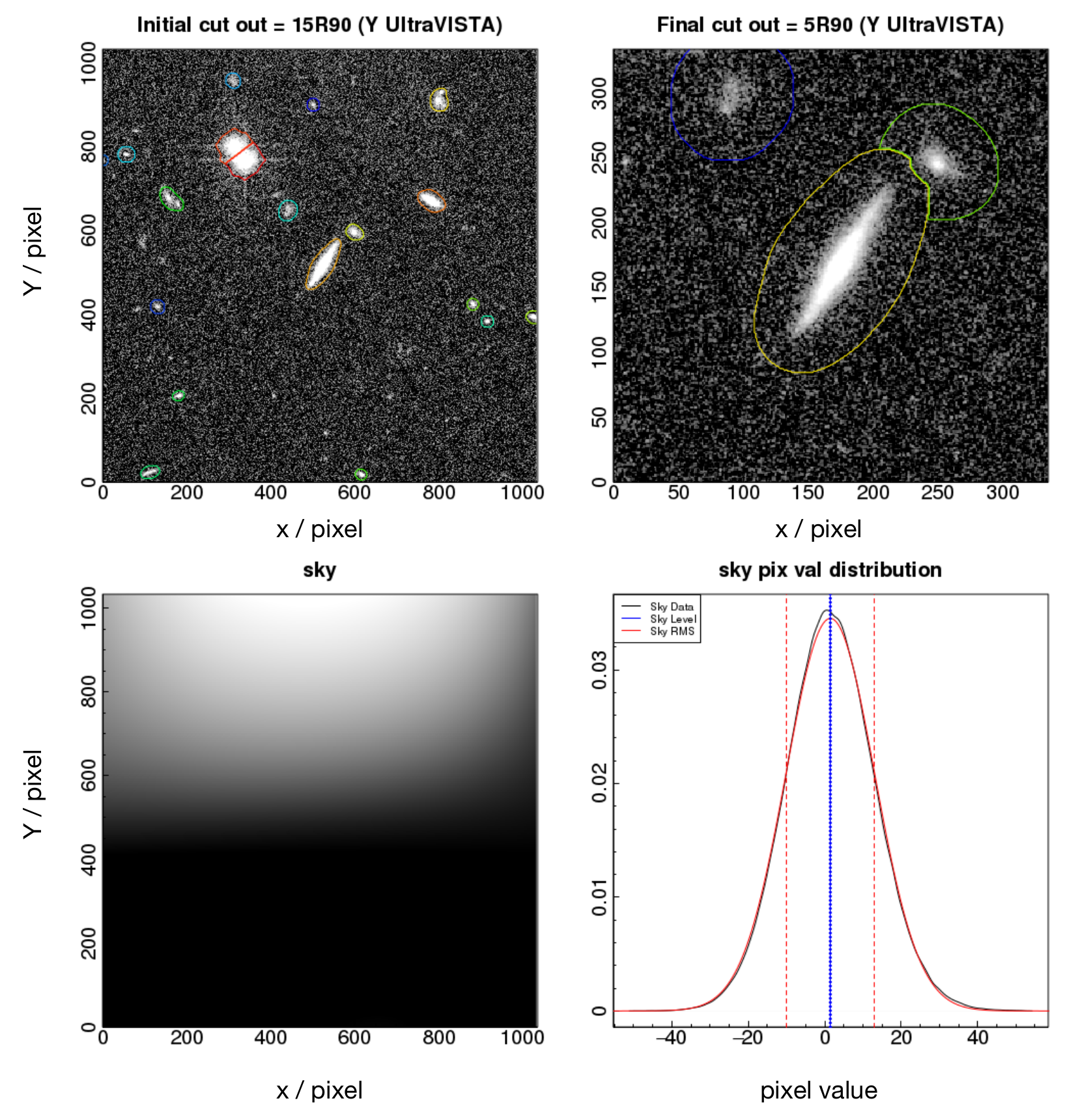}
	\caption{ A set of postage stamps showing the initial cut out of 15$\times$R90 around central galaxy in the top left panel. We perform the source detection and sky estimation on this large cut out. The extracted sources by {\sc ProFound} are over-plotted as isophotal contours. Top right panel displays the final cut out of 5$\times$R90. Bottom left and right panels show the sky image and sky pixel value distribution, respectively. In bottom right panel, the black and red curves represent the distribution of the real sky and sky RMS, respectively. The blue vertical line shows the measured sky level.  }
	\label{fig:postageStams}
    \end{figure}

A robust sky estimation is also crucial in accurately determining source properties in {\sc GRAFit}. For this, we run {\sc ProFound} on our desired image cutout. Table \ref{tab:ProFound_par} shows the {\sc ProFound} settings and argument values that we find work appropriately on our data. To measure the sky and sky RMS, we must use a sufficiently large cutout around the object without creating overly large images. Typically, we can only obtain a reliable sky estimate when $>50 \%$ of the pixels in the cutout region are sky, however, one has to be mindful of extraneous low surface brightness features. As shown in Figure \ref{fig:postageStams}, we therefore use a box car filter with a size scaled to that of the target object and set this to 15$\times$R90 and run {\sc ProFound} on this large cutout (top left panel of Figure \ref{fig:postageStams}). R90 is the radius encompassing 90 percent of the main object's flux. {\sc GRAFit} then sets the box car size for sky estimation as to be 5$\times$R90. The R90 of the objects are taken from the DEVILS input catalog UltraVISTA (\citealt{McCracken12}) Y-band using {\sc ProFound} (see \citealt{Davies18} for more details). Our choice therefore guarantees that more than 50\% of the pixels are real sky pixels. We then use the median and quantile range to estimate a constant sky level as these have been shown to be more stable than the sky mean and standard deviation. For more details on this see the {\sc ProFound} package description\footnote{\href{https://github.com/asgr/ProFound}{https://github.com/asgr/ProFound}}.

Following the sky measurement we perform a final dynamic cutout with 5$\times$R90 (top right panel of Figure \ref{fig:postageStams}) to reduce the number of pixels for a faster run time but {\sc GRAFit} allows the final cutout size to grow if the main source's segmentation touches the outer edges of the cut out (e.g., for extreme inclination objects). An example of this process is shown in Figure \ref{fig:postageStams}. 

The bottom left panel of Figure \ref{fig:postageStams} displays the distribution of the sky pixel values, and the bottom right panel shows the sky and sky RMS in black and red curves, respectively. Therefore, the closer the sky RMS distribution is to the sky distribution, the more accurate sky measurement we achieve. {\sc GRAFit} handles all the above processes using the module \texttt{GRAFitDynamo}.

\begin{table*}
\centering
\caption{{\sc ProFound} setting for main arguments. The rest of massive {\sc ProFound}'s arguments were left as default.}
\begin{adjustbox}{scale = 1}
\begin{tabular}{ll}
\firsthline \firsthline \\

{\sc ProFound} argument & Value \\
\firsthline \\
\texttt{tolerance} & 7 \\
\texttt{sigma} & 7 \\
\texttt{pixcut} & 3 \\
\texttt{skycut} & 1.1 \\
\texttt{smooth} & TRUE \\
\texttt{size} (dilation) & 51 \\
\texttt{ext} & 2 \\
\texttt{box}1 (initial run) & $5\times \mathrm{R90}$ (Y-UltraVista) \\
\texttt{box}2 (final run) & $3\times \mathrm{R90}$ (I-ACS) \\
\texttt{type} & "bicubic" \\
SD of Gaussian Priors & (2,5,0.3,0.3,0.1,30,0.3,$\infty$) \\
(position,mag,Re,bulge-$n$,disk-$n$,angle,axial ratio, boxiness) & \\

\lasthline
\end{tabular}
\end{adjustbox}
\label{tab:ProFound_par}
\end{table*}

\subsection{Other input priors}
\label{subsubsec:priors}

\begin{figure}
	\centering
	\includegraphics[width=\columnwidth]{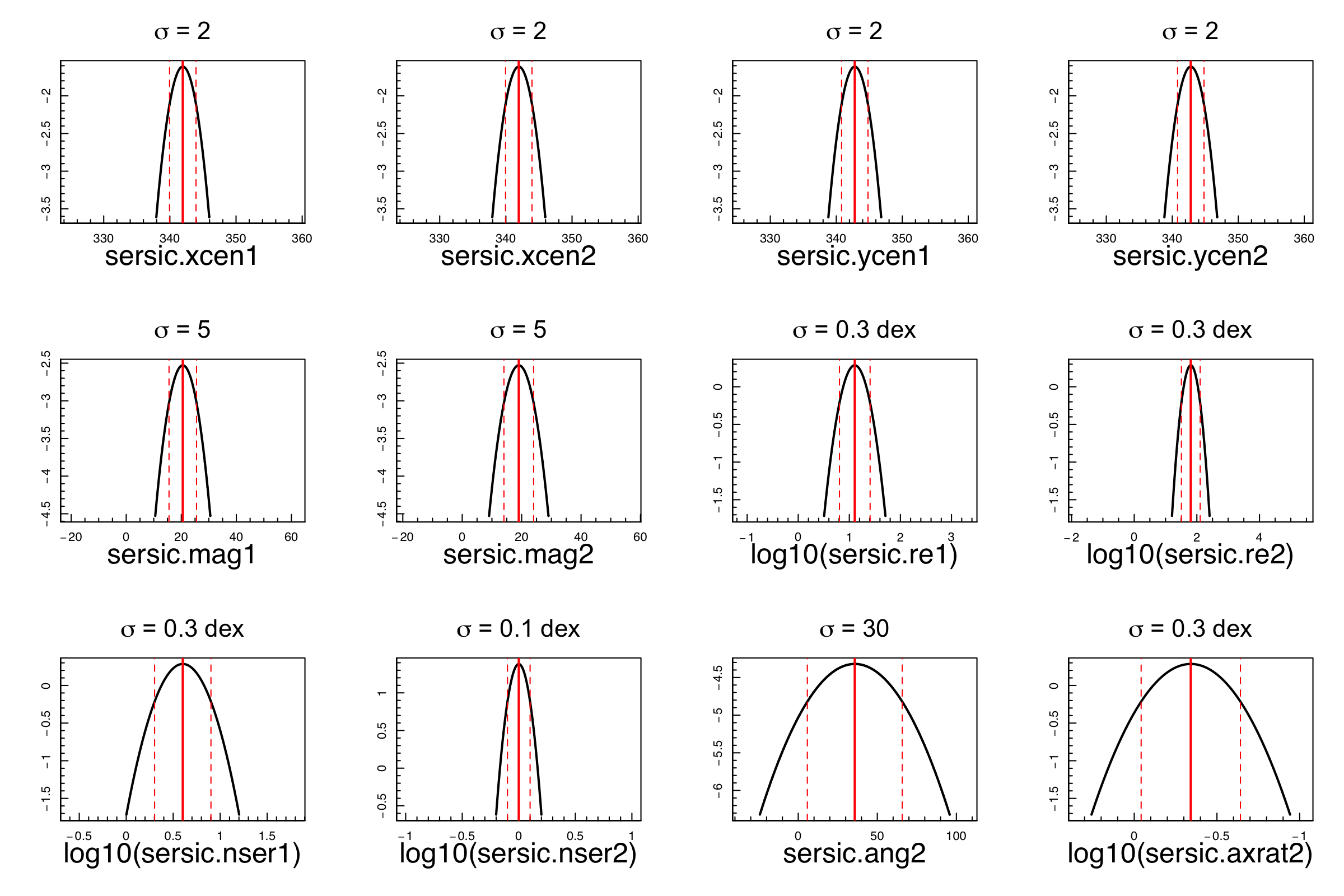}
	\caption{The prior distributions for all the parameters that we fit for a sample galaxy of D10/ACS. We consider Gaussian priors with appropriate mean and standard deviation. The mean values, shown as red solid vertical lines, are generally our initial guesses coming from photometric measurements by {\sc ProFound}. Dashed red vertical lines represent 1$\sigma$ region of the Gaussian distributions. Note that indices 1 and 2 refer to bulge and disk, respectively. Also, re is in the unit of pixel and angle in degree.}
	\label{fig:priorplot}
\end{figure}

 {\sc GRAFit} makes use of the ProFit function \texttt{profitSetupData} to set up all the data in a ProFit-understandable class, and then provides a PSF convolved model image given a set of structural parameters. These include the half-light radius ( $\rm{R}_e$), the S\'ersic index ($n$), the ellipticity ($e$), and the x and y coordinates of the central pixel. 
{\sc GRAFit} also imposes limits on the S\'ersic indices of: 0.5 $\leq$n$\leq$ 1.5 for disks and 0.5 $\leq$n$\leq$ 20 for bulges. 
{\sc ProFit} also accepts prior distributions for each of the parameters. Within {\sc GRAFit} we define Gaussian prior distributions with the mean value of the initial {\sc ProFound} measurements, and standard deviations of $2,5,0.3,0.3,0.1,30,0.3,\infty$ for the central position, magnitude, $\rm{R}_e$, bulge's S\'ersic index, disk's S\'ersic index, position angle, ellipticity and boxiness, respectively. Note $\infty$ corresponds to a flat distribution and S\'ersic indices are fitted in log space, and hence the standard deviation is in dex. Figure \ref{fig:priorplot} displays the prior distribution for a random galaxy in our sample. 
Unlike the parameters' limits (uninformed priors) which allow a wide exploration of the parameter space, the priors are tighter and based on our prior knowledge of galaxy parameters from our {\sc ProFound} photometry. The priors are also not too restrictive and solutions can be found outside the prior range if the data requires it.

{\sc GRAFit} also gives the user the flexibility to choose whether the centres of disk and bulge should be tied together, or allowed to roam to some pixel tolerance, which is sometimes necessary due to disk asymmetry and the presence of dust. In this work, by testing different offsets and performing visual inspections we select the maximum offset of the bulge from the disk to be 11 pixels ($0.33$ arc seconds).  

\subsection{{\sc GRAFit} Outputs} 
\label{sec:GRAFit_out}

\begin{figure}
	\centering
	\includegraphics[width=\columnwidth]{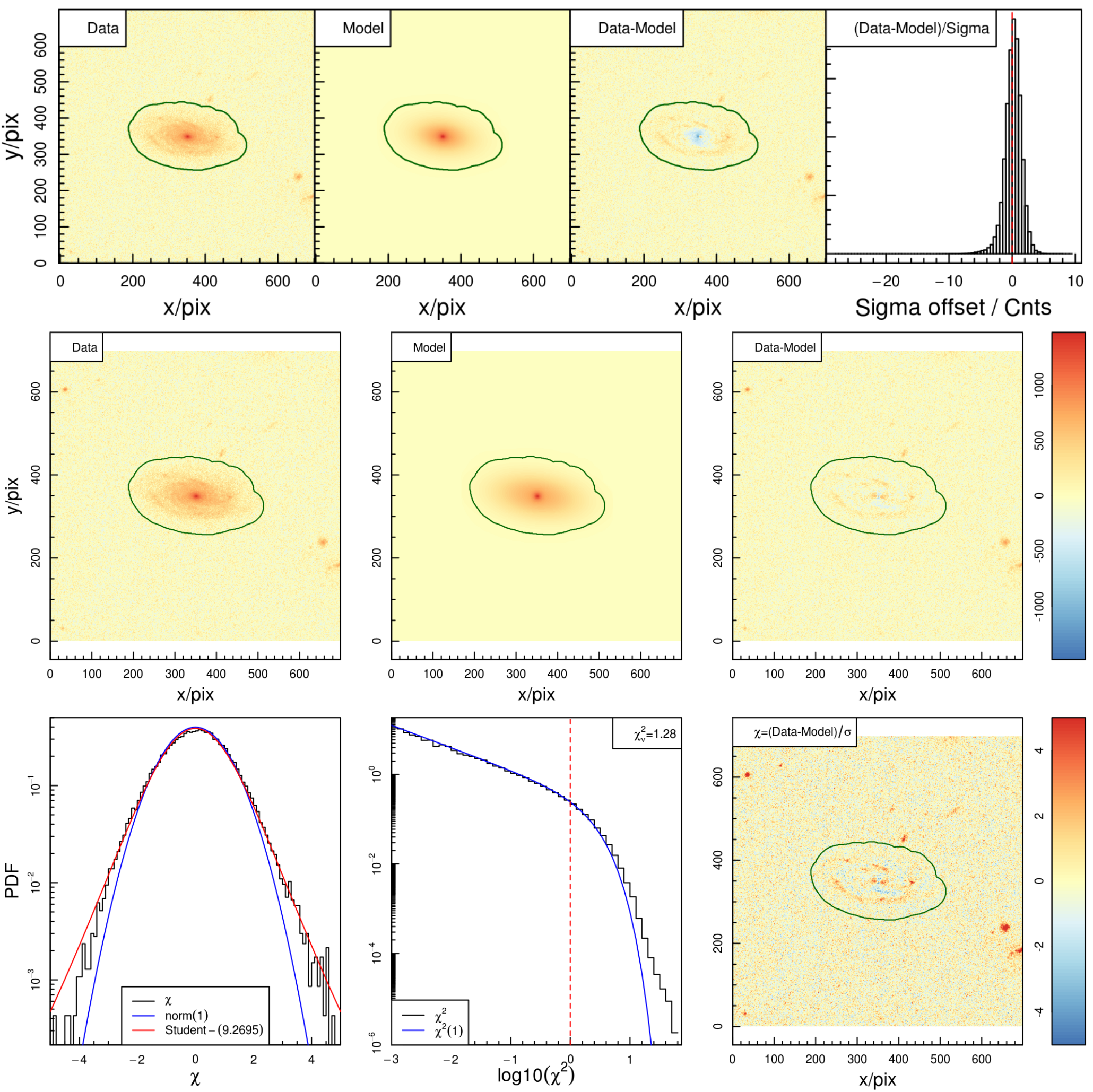}
	\caption{Our double S\'ersic decomposition of galaxy UID = 101494996111806000 highlighting the standard {\sc ProFit} outputs. Top row shows our initial model inferred from initial conditions from {\sc ProFound} photometry. First, second and third panels show the galaxy image (data), initial model and residual (Data-Model), respectively. Forth panel shows the distribution of the residual pixel values ($\chi = \mathrm{Data-Model}/\sigma$), indicating the sigma offset of the model's pixel value from actual image within the segmentation area (green isophotal contours).  
	Middle and lower rows show our final MCMC optimised model. Left, middle and right panels of the middle row show data, final model and residual, respectively. Lower row indicates the residual pixel value distribution (left), the $\chi^2$ distribution with an overlaid $\chi^2$ with one degree of freedom (middle) and the 2D residual pixel map scaled by $\sigma$ (right). The residuals indicate non-smooth structures in this galaxy, in this case spiral arms.  }
	\label{fig:output_init_final}
\end{figure}

The top panels of Figure \ref{fig:output_init_final} show an example of the initial model for galaxy UID = 101494996111806000 at $z = 0.53$. In this Figure, we show the galaxy image (data), our initial model using {\sc ProFound} parameters and the residual. From the residual pixel values and its histogram (right-most panel), we see that our initial parameters provide to first order a relatively good model of the galaxy, where the over- and under-subtracted regions in the residual are mapped with blue and red colours, respectively. 

We then run our initial model through an MCMC optimization process using the CHARM algorithm within the \texttt{LaplacesDemon} package implemented in {\sc R}. We show our final optimized model for the above galaxy in the middle and bottom rows of Figure \ref{fig:output_init_final}. Now, we see that the residual pixel maps (right panels) are significantly improved with most of the non-zero residual pixels highlighting small scale features, such as spiral arms, star-forming clumps and a bar.

\begin{figure}
\centering
\includegraphics[width=0.49\textwidth]{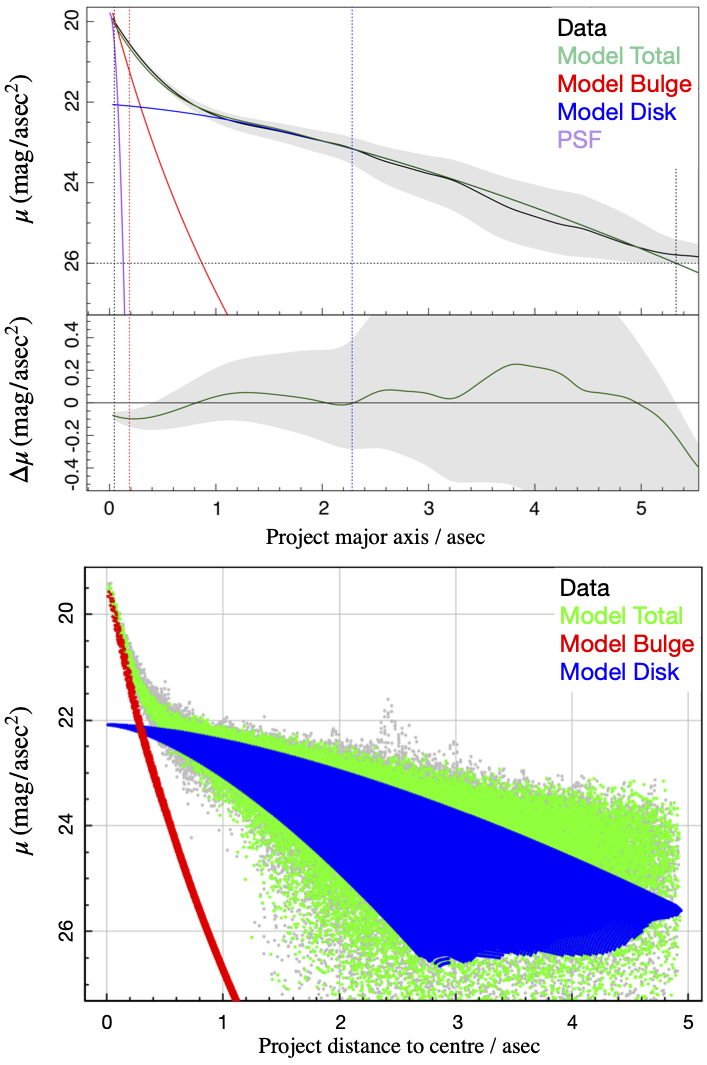}
\label{subfig:stMass_PDF}
\caption{ Top panel: Radial profile of UID = 101494996111806000 highlighting the variation of the surface brightness as a function of projected major axis. Black curve shows the data with its $1\sigma$ region shown as gray area while green, red and blue represent the total, bulge and disk optimized models, respectively. The purple profile shows the PSF. Middle panel: indicates the one-dimensional residual of the image and model profiles. Dashed vertical lines show the half-light-radius, R50, of each profile. Bottom panel: pixel by pixel surface brightness as a function of their distance to the centre of the galaxy.}
\label{fig:1D_pro}
\end{figure}

Figure \ref{fig:1D_pro} shows the collapsed one-dimensional radial profile of our final model for UID = 101494996111806000. The bottom panel shows the surface brightness values of the actual pixels together with total, disk and bulge model pixels. Finally, in Figure \ref{fig:triPlot}, we display the corner plot of the stationary MCMC chain of our double S\'ersic model for UID = 101494996111806000.
Alongside the graphical outputs, {\sc GRAFit} returns a comprehensive catalogue including all the inputs and final model parameters. 

We now run GRAFit over our full sample of $35,803$ galaxies electing to fit three models in each case: 
\begin{enumerate}
\item a single S\'ersic model with free S\'ersic index $n$ ($0.5 < n < 20$).
\item a double S\'ersic model with near exponential disk ($0.5 < n_d < 1.5$) and free S\'ersic index for bulge ($0.5 < n_b < 20$).
\item a double S\'ersic model but with a fixed exponential disk ($n_d = 1$) and free S\'ersic index for bulge ($0.5 < n_b < 20$).
\end{enumerate}
  
\subsection{Parameter convergence test of the MCMC minimisation} 
\label{sec:MCMC}

\begin{figure}
	\centering
	\includegraphics[width = \columnwidth, angle = 0]{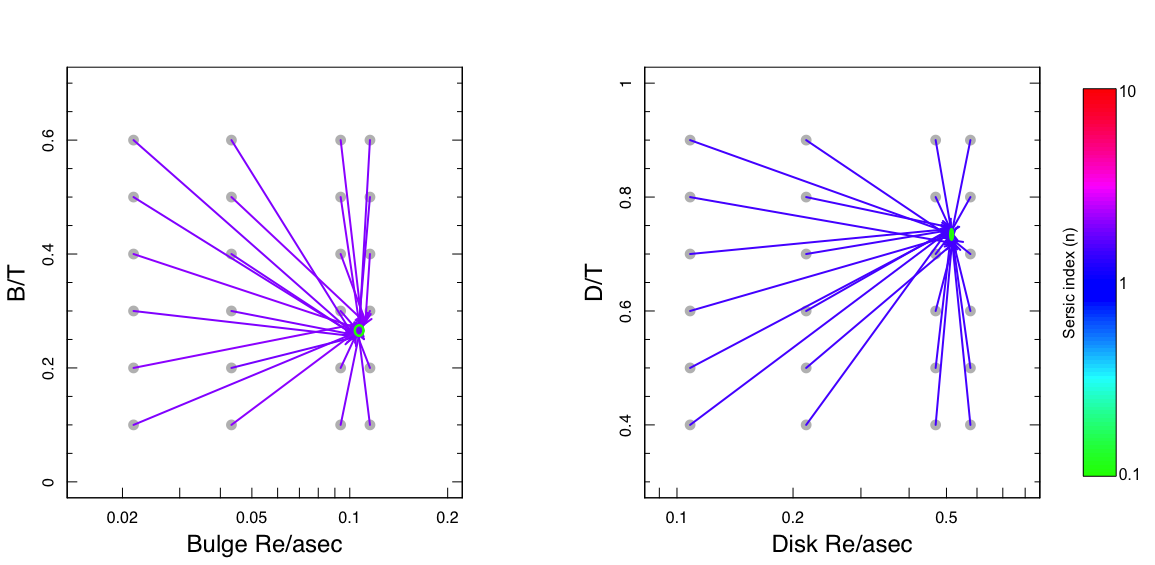}
	\caption{A convergence test of ProFit performance on galaxy UID = 101501367451880000. We start the model from different initial conditions and show that the final model is converged to an identical answer. Left panel shows the relation of B/T with Bulge $\rm{R}_e$ and bulge's S\'ersic index. Arrows indicate the initial and final values colour coded according to the S\'ersic index. Right panel displays the D/T as a function of disk $\rm{R}_e$ and disk's S\'ersic index.}
	\label{fig:Conv_test_HST}
\end{figure}

As a final check of the {\sc ProFit} MCMC approach, we consider the possibility of our algorithm becoming trapped in local minima of the parameter space. To explore this, we address the issue raised by \cite{Lange16} who used {\sc GALFIT3} \citep{Peng10} to fit nearby galaxies from the GAMA survey \citep{Driver11}, and found in many cases a strong dependence of the recovered parameters on the initial input parameters. This was demonstrated by running a grid of input parameters, and in many cases finding convergence was not consistent. We repeated this experiment for a random sample of 20 galaxies across the full $M_*-z$ parameter space, and found that consistently, for a grid of initial conditions, we recover the same solution. An example of this is shown as Figure \ref{fig:Conv_test_HST} for galaxy UID = 101501367451880000 which reflects the results seen for all the galaxies tested in this fashion. We therefore conclude that, our MCMC sampling is able to find unique global solutions for almost 100\% of the sample due to the robustness of these algorithms, and not getting trapped in local minima. To further highlight this point, we specifically explore an extreme failure highlighted by \citealp{Lange16}, i.e., galaxy G32362 (see figure 6 of \citealp{Lange16}), for which we find convergence using {\sc ProFit}, but in this instance applied to the identical Sloan Digital Sky Data as used by \cite{Lange16} (see Appendix \ref{sec:conv_GAMA}).

\subsection{Parameter convergence test on a GAMA galaxy. } \label{sec:conv_GAMA}
To further quantify the accuracy of our MCMC sampling we perform our analysis on the worst case of \cite{Lange16} for which they find the most diverged results, i.e., G32362. Similar to Section \ref{sec:MCMC} we start from a set of initial conditions and find that all fits well converge to a global solution.

\begin{figure}
	\centering
	\includegraphics[width = \columnwidth, angle = 0]{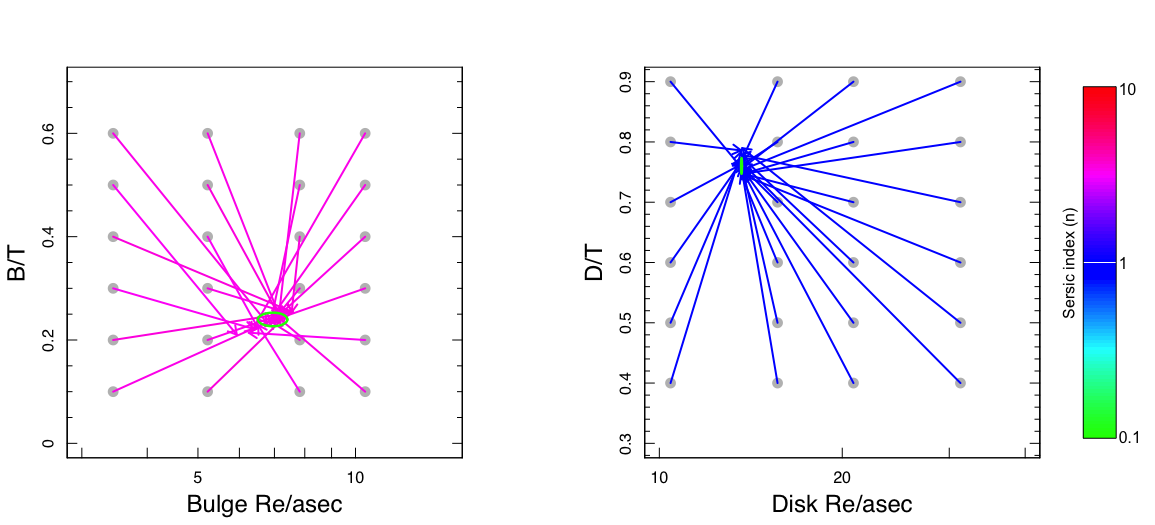}
	\caption{ An MCMC convergence test on GAMA galaxy G32362.}
	\label{fig:Conv_test_GAMA}
\end{figure}

\section{Selection of stars for PSF subtraction} 
\label{sec:star_sel}

\begin{figure}
	\centering
	\includegraphics[width=\linewidth]{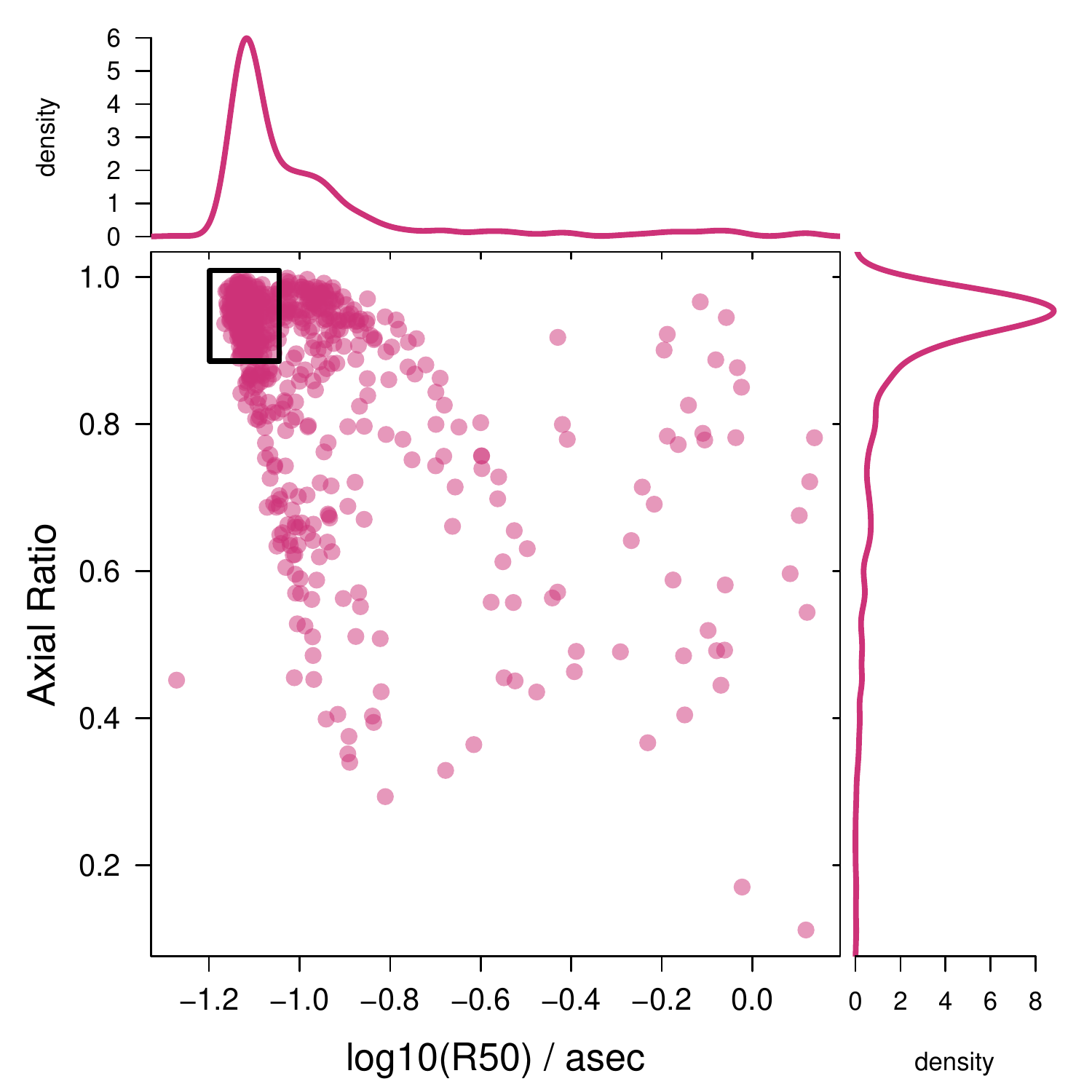}
	\caption{ The main plot shows ellipticity as a function of half light radius (R50) of $\sim700$ random selected unsaturated bright stars. Upper and right marginal histograms show the probability density of R50 and axial ratio, respectively. For star subtraction from PSF, we select 5 random stars in the black rectangle ensuring the selected stars are within the most frequent size (R50 $\sim0.07''$, peak of distribution), and most likely single stars (axial ratio $> 0.9$). 
}
	\label{fig:5star_sel}
\end{figure}

To test the accuracy of our modelled PSF, we select 5 random stars based on their R50 and axial ratio.
The main panel in Figure \ref{fig:5star_sel} shows axial ratio versus R50 for $~700$ stars identified in the mosaic frame. The small black rectangular area is where we randomly select our five stars for which we subtract the corresponding PSFs as presented in Figure \ref{fig:starsub_5star}.

\section{Star-PSF subtraction from a star in raw ACS frames.} 
\label{sec:star_psf_raw}

\begin{figure*} 
	\centering
	\includegraphics[width=\textwidth]{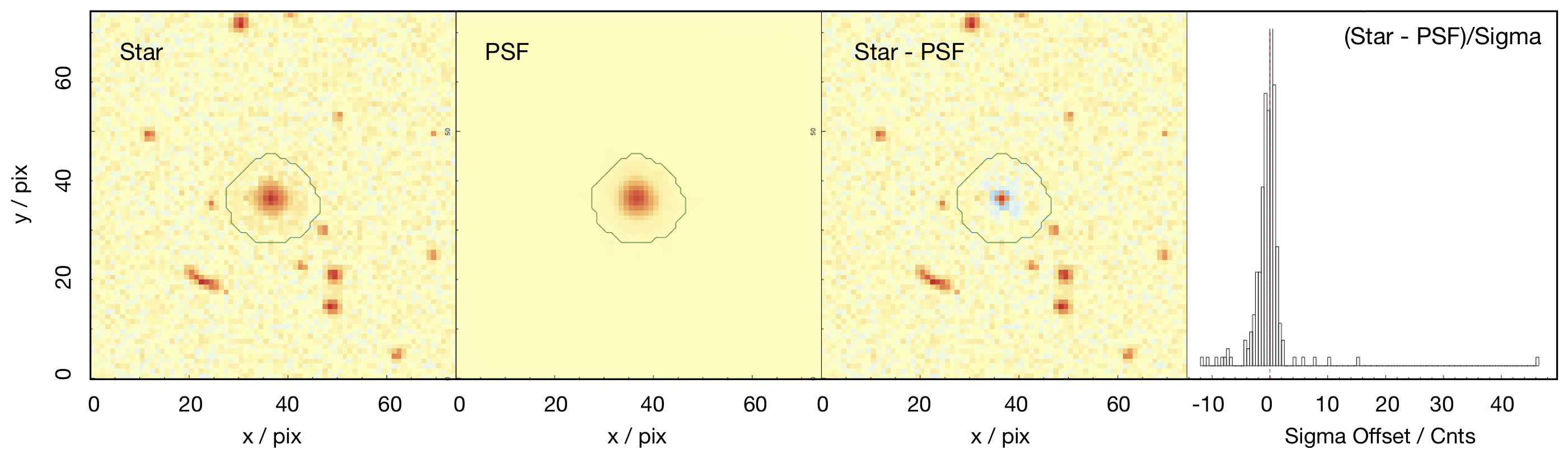}
	\caption{PSF subtraction from a star in a raw HST/ACS frame before cosmic ray rejection, stacking and sub-sampling. Columns are the same as Figure \ref{fig:starsub_5star}. Most of the sources around the star are cosmic rays.}
	\label{fig:starsub_raw}
\end{figure*}

\begin{figure*}
	\centering
	\includegraphics[width=0.8\textwidth]{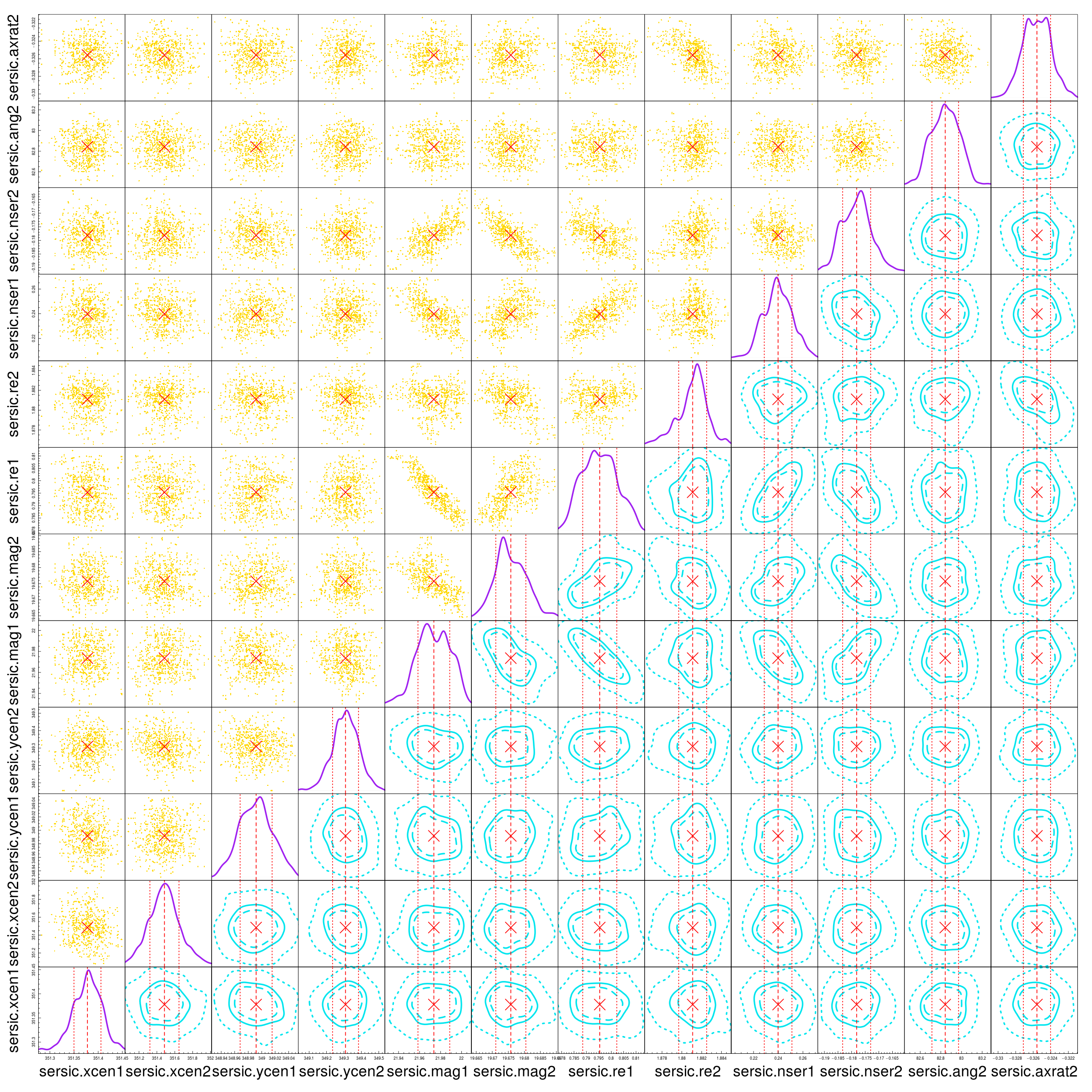}
	\caption{ The corner plot of the stationary MCMC chain of our fitting for UID = 101494996111806000. In this case we fit 12 parameters. The top left corner of the plot shows the scatter sample while the lower right corner shows the contour version of the sample. Dashed, solid and dotted contours contain 50, 68 and 95 per cent of the data, respectively. The diagonal density plots show the one-dimensional marginalized distribution of the sample for each parameter. }
	\label{fig:triPlot}
\end{figure*}

In Section \ref{sec:PSF_test}, we described our method for stacking four PSFs generated by {\sc Tiny Tim}. Here we perform a star subtraction from stars in the HST/ACS raw frames to confirm that the accuracy of our final PSF is not influenced by our stacking procedure. Figure \ref{fig:starsub_raw} shows this process where we subtract a star in raw frame (first panel) from our PSF directly out of {\sc Tiny Tim} (second panel) implying that the residual is still present at the centre (third and fourth panels). 

\section{Stationary MCMC chain} 
\label{sec:MCMC_chain}
The corner plot of the stationary MCMC chain of our double S\'ersic model for UID = 101494996111806000, showing the MCMC chain for each parameter as a scatter plot (top-left corner) alongside their contour version diametrically opposite (i.e., lower-left corner). We also present the diagonal one-dimensional marginalized distribution of the sampling chain.  

\section{The non-LSS-corrected evolution of the SMD} 
\label{sec:MassBuildUp_noLSS}

\begin{figure}
\centering
  \centering
  \includegraphics[width = \columnwidth]{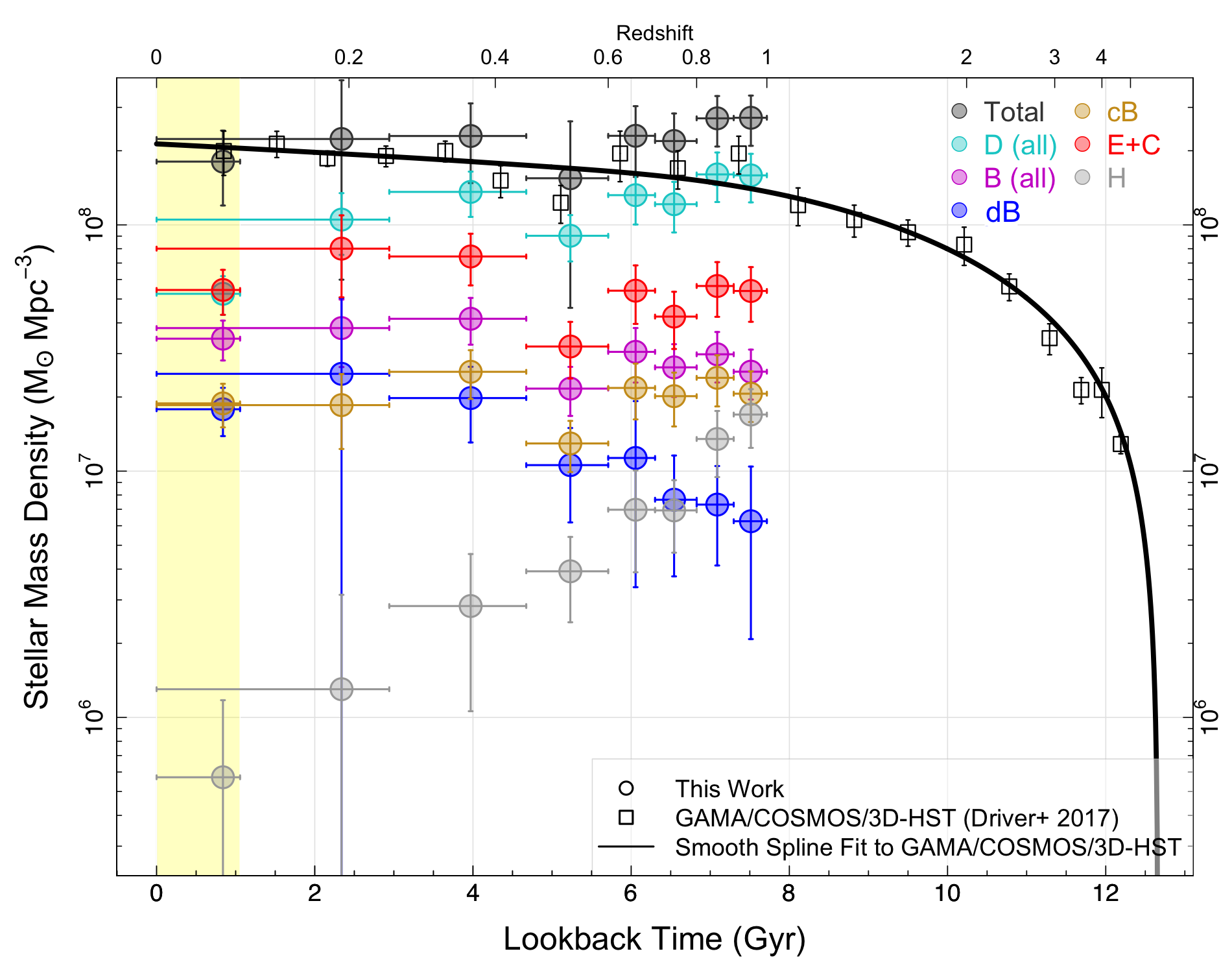}
  \caption{The evolution of the stellar mass density (SMD) of total and morphological types before applying the LSS corrections. Highlighted region shows the epoch covered by the GAMA data.}
  \label{fig:MassBuildUp_noLSS}
\end{figure}

Figure \ref{fig:MassBuildUp_noLSS} shows the evolution of the integrated stellar mass density, $\rho_*$, before we apply our large scale structure corrections. This is to further confirm that the corrections do not derive the overall trends that we find in Figure \ref{fig:MassBuildUp} as explained in Section \ref{sec:rho}.

\label{lastpage}
\end{document}